\documentclass[5p,twocolumn]{elsarticle}

\usepackage{amssymb}
\usepackage{latexsym}

\usepackage{url}
\usepackage{xcolor}
\usepackage{makecell}
\usepackage{tabu, booktabs}
\usepackage{hyperref}
\usepackage{lineno}
\usepackage{bigstrut}
\usepackage{enumitem}
\usepackage{framed,multirow}

\usepackage{amssymb}
\usepackage{latexsym}

\usepackage{threeparttable}
\usepackage{color}
\definecolor{newcolor}{rgb}{.8,.349,.1}
\usepackage{bbm}
\usepackage{mathtools}
\usepackage{bbold}

\begin{document}

\begin{frontmatter}

\title{\textit{SegRap2023}: A Benchmark of Organs-at-Risk and Gross Tumor Volume Segmentation for Radiotherapy Planning of Nasopharyngeal Carcinoma}%

\author[1,2,3]{Xiangde Luo}
\author[1]{Jia Fu}
\author[4]{Yunxin Zhong}
\author[4]{Shuolin Liu}
\author[4]{Bing Han}
\author[5]{Mehdi Astaraki}
\author[6]{Simone Bendazzoli}
\author[5]{Iuliana Toma-Dasu}
\author[7]{Yiwen Ye}
\author[7]{Ziyang Chen}
\author[7]{Yong Xia}
\author[3]{Yanzhou Su}
\author[3]{Jin Ye}
\author[3]{Junjun He}
\author[8]{Zhaohu Xing}
\author[8]{Hongqiu Wang}
\author[8]{Lei Zhu}
\author[9]{Kaixiang Yang}
\author[9]{Xin Fang}
\author[9]{Zhiwei Wang}
\author[10]{Chan Woong Lee}
\author[10]{Sang Joon Park}
\author[11]{Jaehee Chun}
\author[12]{Constantin Ulrich}
\author[12]{Klaus H. Maier-Hein}
\author[13]{Nchongmaje Ndipenoch}
\author[13]{Alina Miron}
\author[13]{Yongmin Li}
\author[14]{Yimeng Zhang}
\author[14]{Yu Chen}
\author[14]{Lu Bai}
\author[15]{Jinlong Huang}
\author[15]{Chengyang An}
\author[15]{Lisheng Wang}
\author[16]{Kaiwen Huang}
\author[16]{Yunqi Gu}
\author[16]{Tao Zhou}
\author[3]{Mu Zhou}
\author[2]{Shichuan Zhang}
\author[2]{Wenjun Liao}
\author[1,3]{Guotai Wang\corref{cores}}\ead{guotai.wang@uestc.edu.cn}
\author[1,3]{Shaoting Zhang\corref{cores}}\ead{Rutgers.shaoting@gmail.com}\cortext[cores]{Corresponding authors}

\address[1]{School of Mechanical and Electrical Engineering, University of Electronic Science and Technology of China, Chengdu, China.}
\address[2]{Department of Radiation Oncology, Sichuan Cancer Hospital \& Institute, Chengdu, China.}
\address[3]{Shanghai AI Lab, Shanghai, China.}
\address[4]{Canon Medical Systems (China) Co. Ltd., Beijing, China.}
\address[5]{Department of Medical Radiation Physics, Stockholm University, Solna, Sweden.}
\address[6]{Department of Biomedical Engineering and Health Systems, KTH, Huddinge, Sweden.}
\address[7]{Northwestern Polytechnical University, Xi’an, China.}
\address[8]{Hong Kong University of Science and Technology (Guangzhou), Guangzhou, China.}
\address[9]{Wuhan National Laboratory for Optoelectronics and with MoE Key Laboratory for Biomedical Photonics.}
\address[10]{Yonsei University College of Medicine, Seoul, South Korea.}
\address[11]{Oncosoft Inc. Seoul, South Korea.}
\address[12]{Division of Medical Image Computing, German Cancer Research Center (DKFZ), Heidelberg, Germany.}
\address[13]{Department of Computer Science, Brunel University London, Uxbridge, United Kingdom.}
\address[14]{MedMind Technology Co. Ltd., Beijing, China.}
\address[15]{Shanghai Jiao Tong University, Shanghai, China.}
\address[16]{Nanjing University of Science and Technology, Nanjing, China.}

\begin{abstract}
Radiation therapy is a primary and effective NasoPharyngeal Carcinoma (NPC) treatment strategy. The precise delineation of Gross Tumor Volumes (GTVs) and Organs-At-Risk (OARs) is crucial in radiation treatment, directly impacting patient prognosis. Previously, the delineation of GTVs and OARs was performed by experienced radiation oncologists. Recently, deep learning has achieved promising results in many medical image segmentation tasks. However, for NPC OARs and GTVs segmentation, few public datasets are available for model development and evaluation. To alleviate this problem, the SegRap2023 challenge was organized in conjunction with MICCAI2023 and presented a large-scale benchmark for OAR and GTV segmentation with 400 Computed Tomography (CT) scans from 200 NPC patients, each with a pair of pre-aligned non-contrast and contrast-enhanced CT scans. The challenge's goal was to segment 45 OARs and 2 GTVs from the paired CT scans. In this paper, we detail the challenge and analyze the solutions of all participants. The average Dice similarity coefficient scores for all submissions ranged from 76.68\% to 86.70\%, and 70.42\% to 73.44\% for OARs and GTVs, respectively. We conclude that the segmentation of large-size OARs is well-addressed, and more efforts are needed for GTVs and small-size or thin-structure OARs. The benchmark will remain publicly available here: \url{https://segrap2023.grand-challenge.org}.

\end{abstract}

\begin{keyword}
Nasopharyngeal carcinoma \sep organ-at-risk\sep gross tumor volume \sep Segmentation
\end{keyword}

\end{frontmatter}

\begin{figure}[t]
    \centering    
    \includegraphics[width=0.99\columnwidth]{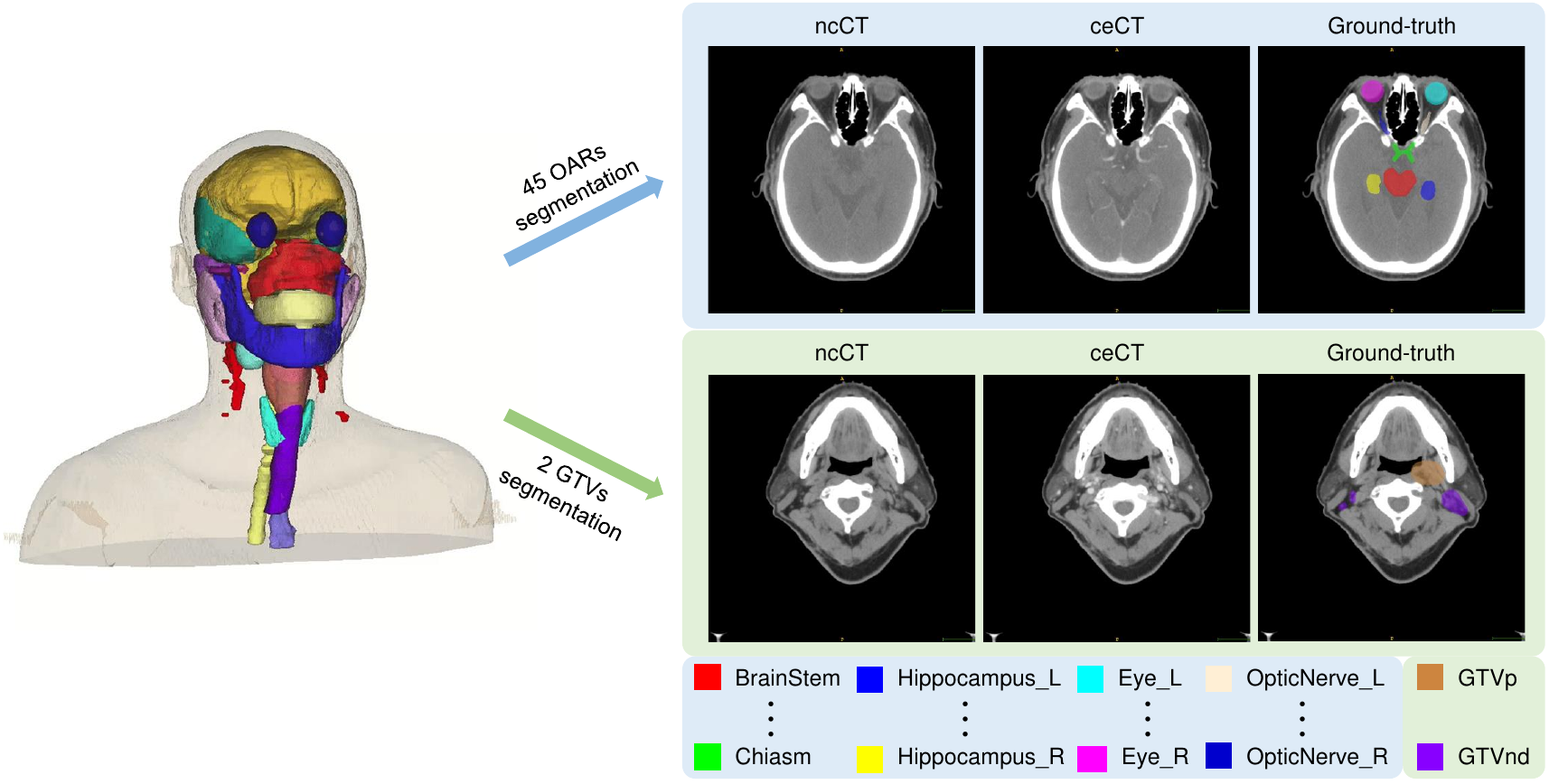}
    \caption{Overview of two sub-tasks in the SegRap2023 challenge.}
    \label{fig:overview}
\end{figure}
\section{Introduction}
\subsection{Clinical background}
Nasopharyngeal carcinoma (NPC), a malignant tumor originating in the nasopharyngeal region, is particularly prevalent in Southeast Asia and North Africa~\citep{lee2015management,chua2016nasopharyngeal,sun2019association}. The primary treatment modality for NPC relies heavily on radiation therapy, especially Intensity-Modulated Radiation Therapy (IMRT)~\citep{xia2000comparison,kam2003intensity}. In IMRT, the accurate delineation of the Gross Tumor Volumes (GTVs) and the surrounding Organs-At-Risk (OARs) is crucial for treatment effectiveness. Accurately identifying the target area is essential to ensure that high doses of radiation precisely cover the tumor while protecting the adjacent normal tissues~\citep{tang2019clinically}. Proper delineation of the GTVs enhances local control rates of the treatment and reduces the risk of recurrence. NPC is located near several vital structures, such as the skull base, internal carotid arteries, and optic nerves~\citep{wang2021guidelines}. Inaccurate delineation may expose these OARs to unnecessarily high doses of radiation, increasing the risk of acute and delayed radiation-induced damage~\citep{lin2019deep}.

\par Accurate delineation of OARs and GTVs is a significant challenge for junior radiation oncologists and automated delineation methods~\citep{chen2021deep}. Firstly, the anatomical structure of the nasopharyngeal region is inherently complex, being near critical organs and neural structures such as the skull base, internal carotid arteries, and optic nerves. This complexity makes the accurate delineation of the target area and OARs extremely challenging and prone to errors~\citep{tang2019clinically}. Additionally, the tumor size, shape, and location vary among NPC patients, coupled with individual anatomical differences, which further complicates the delineation process~\citep{lee2018international}. Moreover, the reliance on the experience and judgment of physicians for delineating the target area and OARs introduces potential variability and subjectivity among different practitioners, potentially leading to inconsistencies in treatment planning. In past clinical practices, the delineation of OARs and GTVs in NPC was predominantly conducted by experienced radiation oncologists. However, according to the clinical treatment guideline, each patient has more than 40 OARs and 2 GTVs need to be delineated accurately~\citep{ye2022comprehensive,guo2020organ}. It requires the radiation oncologists to spend much time performing delineation, increasing the annotator's burden and patient waiting time. It's desirable to develop efficient and accurate automatic segmentation tools to assist and accelerate the clinical delineation workflow and reduce the annotator's burden and patient waiting time.

\subsection{Technical Challenges}
Deep learning-based segmentation methods have shown promising performance on certain medical segmentation datasets, such as abdominal organ segmentation~\citep{ma2021abdomenct,luo2021word,isensee2021nnu,gibson2018automatic,bilic2019liver} and thoracic organ segmentation~\citep{dong2019automatic,feng2019deep}. However, there remains a notable scarcity of studies reporting automatic segmentation tools for OARs and GTVs in NPC that achieve clinically applicable performance on large-scale datasets. The automation of OAR and GTV segmentation remains challenging due to inherent characteristics, including variations in size, shape, and location among NPC patients, compounded by individual anatomical differences and ambiguous boundaries. Moreover, the creation and annotation of a large-scale, high-quality dataset for OAR and GTV segmentation is a resource-intensive process, demanding both expertise and time to generate accurate delineations. Consequently, utilizing a large-scale dataset for algorithmic investigation to address these clinical inherent characteristics is challenging to implement.

\par Recently, few studies have reported in detail the segmentation results of GTVs and OARs of NPC~\citep{liu2021dilated,lin2019deep,luo2023deep,luo2021efficient,liao2022automatic,ye2022comprehensive,guo2020organ,shi2022deep,tang2019clinically}. Most of them only focused on the segmentation of part of the OARs or the GTVs of head and neck cancers. In addition, few works investigated the model segmentation performance when using multi-modality data as model inputs, such as no-contrast or contrast-enhancement CT scans~\citep{wang2020automated,oreiller2022head}. They may lead to the automatic segmentation methods performing poor performance and generalising on the real clinical applications. Therefore, there is a need to build a large-scale benchmark with exhausted and high-quality annotations and multiple modalities for comprehensive evaluation. 

\begin{table*}[t]
    \centering
    \caption{Summary of several publicly available organ-at-risk segmentation Computed Tomography (CT) datasets. ceCT is the contrast-enhanced Computed Tomography. ncCT means the non-contrast Computed Tomography.}
    \scalebox{0.8}{\begin{tabu}{c|c|c|c|c|c}
    \hline
    Dataset& Modality& No. of categories& Scans (Training/Testing)& Year& Link\\
    \hline
    PDDCA & ncCT & 9 OARs & 48 (33/15) & 2015 & \url{www.imagenglab.com/newsite/pddca} \\
    HNC & ncCT & 28 OARs & 35 (18/17) & 2015 &\url{https://wiki.cancerimagingarchive.net/x/xwxp} \\
    HNPETCT & ncCT & 28 OARs & 105 (52/53) & 2017 & \url{https://doi.org/10.7937/K9/TCIA.2017.8oje5q00} \\
    StrucSeg2019 & ncCT & 22 OARs & 60 (50/10) & 2019 & \url{https://structseg2019.grand-challenge.org} \\
    HaN-Seg2023 & ncCT and MRI & 30 OARs & 56 (42/16) & 2023 & \url{https://han-seg2023.grand-challenge.org} \\
    \hline
    \textbf{SegRap2023} & ncCT and ceCT & \textbf{45 OARs} & \textbf{200 (140/60)} & \textbf{2023} & \url{https://segrap2023.grand-challenge.org} \\
    \hline
    \end{tabu}}
    \label{tab:datasets_summary}
\end{table*}

\subsection{Contribution}
To comprehensively evaluate the performance of state-of-the-art (SOTA) algorithms for automatic OARs and GTVs segmentation in NPC RT planning, we organized the SegRap2023 challenge in conjunction with MICCAI2023. The key contributions of this work can be summarized as three-fold. First, we built the first large-scale public dataset of 200 NPC patients where each patient has pre-aligned non-contrast and contrast-enhanced CT scans with ground truth of 45 OARs and 2 GTVs. Second, we presented the SegRap2023 challenge set-up and organized it via the grand challenge platform~\footnote{\url{https://segrap2023.grand-challenge.org/}}. There are a total of 387 teams registered during the model development phase. In the final evaluation phase, 12 and 11 teams submitted their solutions for the OARs and GTVs tasks, respectively. Third, we evaluated, ranked, summarized, analyzed and discussed the results of all submissions. We believe this dataset and challenge can bring benefits to the whole community.
\par The paper summarises the SegRap2023 challenge and is organized as follows. Section~\ref{sec:set2} reviews the existing datasets and methods for this problem. Then, Section~\ref{sec:set3} presents the details of the challenge in the aspects of data collection and annotation, challenge organization and evaluation. Details of all submitted methods are illustrated in Section~\ref{sec:set4}. Afterwards, the analysis and description of the results are presented in Section~\ref{sec:set5}. Finally, we conclude and discuss the SegRap2023 challenge in Section~\ref{sec:set6} and ~\ref{sec:set7}, respectively. 

\section{Realted Works}\label{sec:set2}
\subsection{OAR segmentation in head and neck}
\subsubsection{Benchmarks and datasets}
OAR segmentation plays an irreplaceable role in head and neck cancer radiation therapy planning. Developing an accurate and robust automatic segmentation model always relies on large-scale annotated datasets. However, there are very limited publicly available datasets as collecting and annotating a large-scale dataset is very challenging and expensive and data privacy protection~\citep{wang2023medfmc,kiryati2021dataset,simpson2019large}. Table~\ref{tab:datasets_summary} summarized several public datasets for OAR segmentation in the head and neck region. PDDCA~\citep{raudaschl2017evaluation} provided 48 CT scans with 9 OARs annotations for the Head and Neck Auto Segmentation MICCAI Challenge (2015). After the challenge, the PDDCA dataset was widely used as a benchmark for head and neck OAR segmentation model development and evaluation. HNC
~\citep{ang2014randomized} and HNPETCT
~\citep{vallieres2017radiomics} consists of CT scans selected from two public datasets, all of the patients were diagnosed with head and neck cancer. 

\par In the following work, ~\cite{tang2019clinically} selected 35 CT scans from HNC and 105 CT scans from HNPETCT for further annotation and released all masks for public research, where each patient has 28 OARs labels. StructSeg2019~\footnote{\url{https://structseg2019.grand-challenge.org}} provided 60 CT scans with 22 OARs and GTVp annotations of nasopharyngeal cancer patients for automatic structure segmentation methods development and evaluation in radiotherapy planning. More recently, HaN-Seg2023
~\citep{podobnik2023han} organized a head and neck organ-at-risk segmentation from CT and Magnetic Resonance Imaging (MRI) challenge conjoint with MICCAI2023. The HaN-Seg2023 consists of 56 patients with head and neck cancer and each patient has a CT and a T1-weighted MRI scan and a reference annotation with 30 OARs. 

\par Although these datasets have facilitated the methods research of head and neck OAR segmentation in the community, they may be still not enough to develop clinically applicable segmentation tools and provide comprehensive evaluations due to the small scale and lack of exhaustive annotations. In other medical image segmentation tasks, such as abdominal organ segmentation~\citep{ma2021abdomenct,luo2021word,gibson2018automatic,bilic2019liver}, many large-scale datasets can be used for foundation model development and evaluation and also advance the automatic segmentation methods to be applied in clinical practice~\citep{chen2021deep,kirillov2023segment,huang2023stu,wang2023sam,wang2023mis}. Therefore, for the head and neck OAR segmentation, it is desirable to build a large-scale dataset and benchmark (SegRap2023) to boost technical improvements and clinical application development.


\subsubsection{SOTA OAR segmentation methods in head and neck}

\par Recently, deep learning-based segmentation methods have shown superiority in producing more accurate and robust than previous atlas-based counterparts~\citep{tang2019clinically,kosmin2019rapid,chen2021deep}. Here, we reviewed several classical and famous works about the head and neck OAR segmentation methods. FocusNetV2~\citep{gao2021focusnetv2} presented a two-stage framework to locate and segment OARs progressively by combining the multi-scale convolutional neural network and a shape adversarial constraint. It was evaluated on a large-scale private nasopharyngeal cancer dataset with 1164 CT scans and 22 OARs and the public PDDCA dataset and showed a mean dice score of 82.98\% and 84.50\%, respectively. UaNet~\citep{tang2019clinically} proposed a combination framework to detect OARs and segment them step-by-step, which was trained on a private dataset with 215 CT scans and 28 OARs and tested on 100 CT scans with a mean dice score of 78.34\%. 
\par Recently,~\cite{guo2020organ} and ~\cite{ye2022comprehensive} developed an auto-contouring system (SOARS) by combining the neural architecture search strategy and an organ-level stratification learning. The proposed SOARS was trained on an internal private dataset with 176 CT scans and 42 OARs and independently evaluated on several external cohorts with a total of 1327 CT scans with mean dice scores ranging from 74.80\% to 78.00\%. Additionally,~\cite{he2024statistical} introduced a statistical deformation model-based data augmentation strategy to boost the training set's diversity and realism and further advance the model performance. The proposed was trained and tested on the HNPETCT dataset and achieved a mean dice score of 79.49\%. Based on the above results, we can find that the results of OAR segmentation are significantly different, especially the results on private datasets were higher than on the public datasets~\citep{zhu2019anatomynet,tang2019clinically,ye2022comprehensive,gao2021focusnetv2,he2024statistical,chen2021deep}. Therefore, building a large-scale public benchmark for a fair comparison across multiple SOTA methods is essential.

\subsection{NPC GTV segmentation}
\subsubsection{Benchmarks and datasets}
In this work, the GTVs of NPC consist of the primary gross tumor volume (GTVp) and the lymph node gross tumor volume (GTVnd). According to the clinical statistics, accurate GTVs delineation can improve the patient's 5-year survival ratio and reduce the risk of recurrence and distant metastasis~\citep{burnet2004defining,jin2022towards}. In addition, the accurate GTVs contours can provide a good reference for clinical target volume delineation to avoid under-treatment~\citep{jin2022towards}. In the clinical flow, the GTVs delineation is mostly done on the ncCT as the dose generation relies on accurate delineations and the radiodensity intensity (Hounsfield unit, simplified as HU) of the ncCT scan~\citep{njeh2008tumor,sahbaee2017effect}. Due to the unclear boundary between GTVs and other soft tissues, radiation oncologists usually require other modality images for complementary guidelines to perform GTVs contours, such as ceCT, MRI, fluorodeoxyglucose-positron emission tomography (FDG-PET), etc~\citep{lee2018international,liao2022automatic}. Because of the challenges of imaging, developing powerful automatic GTV segmentation models usually needs more comprehensive and high-quality datasets. 

\par For the GTVs of head and neck cancer segmentation, the public dataset HECKTOR~\footnote{\url{https://hecktor.grand-challenge.org}} was widely used for model development and evaluation. HECKTOR~\citep{oreiller2022head} challenge has been organized in conjunction with MICCAI in recent three years, which aims to encourage all participants to develop cut-edge GTVp and GTVnd segmentation models from CT and FDG-PET scans. The total number of patients increased from 254 patients just with GTVp annotation in HECKTOR2020 to more than 880 patients with both GTVp and GTVnd annotations in HECKTOR2022. The HECKTOR challenge has released the largest scale dataset for research on the GTVs of head and neck cancer segmentation. But for the GTVs of NPC segmentation, there is a very small dataset, StructSeg2019~\footnote{\url{https://structseg2019.grand-challenge.org}}, that can be accessed. The StructSeg2019 provided 60 nasopharyngeal carcinoma patients' CT scans and each patient had a GTVp annotation. Although the HECKTOR challenge provided a large-scale dataset for GTVp and GTVnd segmentation, they focused on head and neck cancer rather than nasopharyngeal carcinoma, so the SegRap2023 is still an important dataset for the GTVp and GTVnd of NPC segmentation.


\subsubsection{SOTA NPC GTV segmentation methods}
Different from OAR segmentation, GTV segmentation has traditionally been conducted by experienced radiation oncologists in clinical practice. This is attributed to the intricate nature of GTVs structures and their significant correlation with prognosis. Moreover, the scarcity of publicly available datasets has been a notable challenge in the field. Many prior studies have reported GTV segmentation outcomes based on privately collected datasets, posing difficulties for both reproducibility and equitable comparisons in the whole community.~\cite{li2019tumor} trained a basic U-Net~\citep{ronneberger2015u} to segment GTVp and GTVnd using a large-scale private dataset with 502 CT scans and achieved a mean dice of 65.86\% and 74.00\% for GTVp and GTVnd, respectively.~\cite{lin2019deep} developed a 3D segmentation model on an MRI dataset with 1021 patients to segment the GTVp and reported the performance with a mean dice score of 79.00\%.~\cite{mei2021automatic} proposed a 2.5D segmentation network with multi-scale and spatial attention to segment GTVp from CT scans and won second place in the StructSeg2019 challenge with a mean dice of 65.66\%.
\par In addition,~\cite{luo2021efficient} proposed a multi-scale consistency-based semi-supervised learning framework to utilize the unlabeled data for GTVp and GTVnd segmentation performance improvement and further demonstrated the applicable in the clinical delineation flow on a private MRI dataset with 258 patients~\citep{liao2022automatic}, where the mean dice scores of GTVp and GTVnd were 83.00\% and 80.00\%, respectively. Recently,~\cite{luo2023deep} did a comprehensive evaluation of GTVp segmentation using a total number of 1057 patients from 5 hospitals and achieved results with a mean dice score of 88.00\% on the multi-center testing cohorts. According to these observations, it can be noted that there is a substantial variation in segmentation results across different datasets. Meanwhile, we can also find that the results based on MRI were significantly superior to those using CT scans. However, the current radiotherapy treatment method is mostly based on CT scans, so accurately contouring the GTVs of NPC from CT scans is still challenging and urgent~\citep{sahbaee2017effect}.

\section{SegRap2023 challenge setup}\label{sec:set3}
\subsection{Challenge Overview}

To evaluate existing and new methods for OAR and GTV segmentation, we organised the SegRap2023 challenge in conjunction with MICCAI2023. The challenge released 400 CT scans from 200 NPC patients where each patient has a pre-aligned pair of ncCT and ceCT scans, to encourage the development of cut-edge and clinically applicable models. Fig.~\ref{fig:overview} shows an overview of the SegRap2023 challenge. The challenge consists of two sub-tasks. The first one (Task01) is to segment 45 OARs, and the second task (Task02) is to segment 2 GTVs.


\begin{table}[t]
    \centering
    \caption{Clinical characteristics of the SegRap2023 training, validation and testing sets. $^*$ means the values are presented as median (range).}
    \scalebox{0.74}{\begin{tabu}{llll}
    \hline
    Characteristics& Training (n=120) & Validation (n=20)& Testing (n=60)\\
    \hline
    Sex & & \\
    ~~~~~~Male& 81 (67.5\%) & 12 (60\%) & 37 (61.7\%) \\
    ~~~~~~Female&39 (32.5\%)& 8 (40\%) & 23 (38.3\%)  \\
    Age$^*$ (years)& 48 (22-74) & 50 (36-69) & 47 (22-70) \\
    T stage& & \\
    ~~~~~~T1& 12 (10\%) & 2 (10\%) & 7 (11.7\%) \\
    ~~~~~~T2& 27 (22.5\%) & 5 (25\%) & 13 (21.7\%) \\
    ~~~~~~T3& 62 (51.7\%) & 11 (55\%) & 32 (53.3\%) \\
    ~~~~~~T4& 19 (15.8\%) & 2 (10\%) & 8 (13.3\%) \\
    N stage& & \\
    ~~~~~~N0& 10 (8.3\%) & 1 (5\%) & 4 (6.7\%) \\
    ~~~~~~N1& 24 (20\%) & 3 (15\%) & 11 (18.3\%) \\
    ~~~~~~N2& 54 (45\%)& 11 (55\%) & 31 (51.7\%)\\
    ~~~~~~N3& 32 (26.7\%) & 4 (20\%) & 14 (23.3\%) \\
    Resolution (mm) & & & \\
    ~~~~~~Inter-plane & 3.0 & 3.0 & 3.0 \\
    ~~~~~~Intra-plane$^*$ & 0.55 (0.43-1.13)& 0.54 (0.49-0.60) & 0.59 (0.45-1.34) \\
    
\hline
    \end{tabu}}
    \label{tab:clinical_info_summary}
\end{table}

\subsection{Data description}
The SegRap2023 dataset consists of 200 newly treated NPC patients from Sichuan Cancer Hospital $\&$ Institute, Sichuan Cancer Center, Chengdu, China. The data acquisition was approved by the Sichuan Cancer Hospital $\&$ Institute ethics board and the private information of each patient has been anonymized. Each patient has a ncCT scan and a ceCT scan. All CT scans are collected by Siemens CT scanners with the following scanning conditions: bulb voltage, 120 kV; current, 300 mA; scan thickness, 3.0 mm; resolution, 1024 $\times$ 1024 or 512 $\times$ 512; injected contrast agent, iohexol (volume, 60$-$80 mL; rate, 2 mL/s; delay, 50 s). Table~\ref{tab:clinical_info_summary} lists the clinical characteristics of the training, validation and testing sets. It can be found that there is a similar distribution of clinical characteristics in the training, validation and testing sets (age, sex, T, N, M stages and inter- or intra-plane spacings).

\par To build the dataset, we retrospectively collected 200 newly treated NPC patients from December 2018 to December 2019. The inclusion criteria were defined as (a) Patients who were histologically confirmed as NPC in the M.D. S.C. Zhang treatment group; (b) The treatment strategy included radiotherapy; (c) The radiotherapy planning had ncCT and ceCT scans and 45 OARs and 2 GTVs annotations; (d) Patients who are live and not recurrent until December 2022. The initial contours of OARs and GTVs were delineated by S.C. Zhang (MD, with more than twenty years of experience in oncology radiation therapy) and their team using the commercial radiotherapy planning software MIM Software~\footnote{\url{https://www.mimsoftware.com}}. Note that, during the initial delineation stage, the radiation oncologists referred to other images (MRI, PET) for clear contours, especially for the GTVs delineation. To build the high-quality dataset, we further invited W. Liao (MD, with ten years of experience in oncology radiation therapy) and S.C. Zhang to check and refine these annotations using ITK-SNAP~\citep{yushkevich2006user}. These annotated OARs and GTVs are the Brain, BrainStem, Chiasm, Cochlea left (Cochlea\_L), Cochlea right (Cochlea\_R), Esophagus, Eustachian tube bone left (ETbone\_L), Esophagus, Eustachian tube bone right (ETbone\_R), Eye left (Eye\_L), Eye right (Eye\_R), Hippocampus left (Hippocampus\_L), Hippocampus right(Hippocampus\_R), Internal auditory canal left (IAC\_L), Internal auditory canal right (IAC\_R), Larynx, Larynx glottic (Larynx\_Glottic), Larynx supraglottic (Larynx\_Supraglot), Lens left  (Len\_L), Lens right (Len\_R), Mandible left (Mandible\_L), Mandible right (Mandible\_R), Mastoid left (Mastoid\_L), Mastoid right (Mastoid\_R), Middle Ear left (MiddleEar\_L), Middle ear right (MiddleEar\_R), Optic nerve left (OpticNerve\_L), Optic nerve right (OpticNerve\_R), Oral cavity, Parotid left (Parotid\_L), Parotid right (Parotid\_R), Pharyngeal constrictor muscle (PharynxCont), Pituitary, SpinalCord, Submandibular left (Submandibular\_L), Submandibular right (Submandibular\_R), Temporal lobe left (TemporalLobe\_L), Temporal lobe right (TemporalLobe\_R), Thyroid, Temporomandibular joint left (TMjoint\_L), Temporomandibular joint right (TMjoint\_R), Trachea, Tympanic cavity left (TympanicCavity\_L), Tympanic cavity right (TympanicCavity\_R), Vestibular semicircular canal left (VestibulSemi\_L), Vestibular semicircular canal right (VestibulSemi\_R) and primary gross tumor volume (GTVp) and lymph node grooss tumor volume (GTVnd). Afterwards, we provided an official data split including training, validation and testing sets with 120, 20 and 60 patients respectively according to clinical characteristics, as detailed in Table~\ref{tab:clinical_info_summary}.

\subsection{Evaluation and rank strategies}
The challenge employed two widely used evaluation metrics to measure the performance of each submission: (1) a region overlap-based metric, Dice Similarity Coefficient (DSC, range from 0 to 1) and (2) a distance-aware metric, Normalized Surface Dice (NSD, range from 0 to 1)~\citep{nikolov2021clinically}. If a submission has some missing target OARs or GTVs on test cases, the corresponding DSC and NSD will be set to 0. Then, we calculated the average DSC and NSD of each OAR or GTV across all testing patients, respectively. Afterwards, we applied the ranking strategy~\citep{bakas2018identifying} to obtain each OAR or GTV score across all participant teams and each team has 45$~\times~$2 or 2$~\times~$2 ranking scores for OAR or GTV segmentation tasks. Finally, we employed the average OARs or GTVs ranking scores of each team for the final ranking.

\subsection{Challenge setup}
In the SegRap2023 challenge, we designed two sub-tasks to evaluate the 45 OARs (Task01) segmentation and GTVs (Task02) segmentation performance, respectively. The challenge consists of three phases (training, validation and testing) and all of them were hosted in the grand challenge platform~\footnote{\url{https://grand-challenge.org}}. During the training stage, all participants can access the training set by signing and sending back an end-user agreement file. After the challenge, the training set can be accessed without any requirement. Afterwards, an automatic evaluation Docker container with two public Python packages (\textit{Evalutils}~\footnote{\url{https://evalutils.readthedocs.io/en/latest}} and \textit{MedPy}~\footnote{\url{https://loli.github.io/medpy}}) was running online to evaluate each participant's submitted algorithm Docker. The validation phase is open from July 10th, 2023 to August 20th, 2023 and each team was allowed to submit 5 times. In the final testing phase, due to the testing set is not accessible~\citep{maier2020bias}, each team was required to submit their solution Docker container for evaluation and ranking. We provided a tutorial~\footnote{\url{https://github.com/HiLab-git/SegRap2023}} to containerise the algorithm with Docker. Each team was only allowed to successfully submit the Docker container once. All submitted Docker containers were run on the grand challenge platform upon they were submitted successfully and then calculated their final ranking score. Finally, the final leaderboard was announced in the MICCAI2023 challenge event after the organization team carefully reviewed and excluded the teams without submitting their technical reports.

\section{Overview of participating methods}\label{sec:set4}
A total of 387 teams registered for the SegRap2023 Challenge, allowing them to download the training data. During the testing phase, 12 teams submitted containerized algorithms for OAR segmentation, while 11 teams submitted containerized algorithms for GTV segmentation. In this section, we summarize the methods employed by the participating teams (two teams were excluded due to the lack of their technical report). More details and references can be found at: \url{https://github.com/HiLab-git/SegRap2023}.

\begin{table*}[t]
  \centering
  \caption{Rankings of methods in DSC/NSD scores for OAR segmentation.}
  \resizebox{0.9\textwidth}{!}{
    \begin{tabular}{ccccccccccc}
    \toprule
    Team & Y. Zhong \textit{et al.} & Y. Ye \textit{et al.} & Y. Su \textit{et al.} & K. Yang \textit{et al.} & C. Lee \textit{et al.} & M. Astaraki \textit{et al.} & Z. Xing \textit{et al.} & Y. Zhang \textit{et al.} & J. Huang \textit{et al.} & K. Huang \textit{et al.} \\
    \bottomrule
    Brain & 4/3   & 2/2   & 1/1   & 3/4   & 7/7   & 5/5   & 8/6   & 9/8   & 6/9   & 10/10 \\
    BrainStem & 1/1   & 3/3   & 5/4   & 7/6   & 10/9  & 8/8   & 2/2   & 4/7   & 6/5   & 9/10 \\
    Chiasm & 4/3   & 2/1   & 8/7   & 7/6   & 3/8   & 6/5   & 5/4   & 1/10  & 10/2  & 9/9 \\
    Cochlea\_L & 1/1   & 3/3   & 2/2   & 6/5   & 4/4   & 5/6   & 9/9   & 8/7   & 7/8   & 10/10 \\
    Cochlea\_R & 1/1   & 3/3   & 2/2   & 6/4   & 5/6   & 4/5   & 9/8   & 8/7   & 7/9   & 10/10 \\
    Esophagus & 2/2   & 4/4   & 1/1   & 3/3   & 5/5   & 6/6   & 8/7   & 9/8   & 7/9   & 10/10 \\
    ETbone\_L & 1/1   & 3/3   & 2/2   & 7/5   & 4/4   & 6/6   & 8/8   & 5/9   & 9/7   & 10/10 \\
    ETbone\_R & 1/1   & 3/2   & 2/3   & 6/4   & 5/6   & 4/5   & 9/9   & 8/7   & 7/8   & 10/10 \\
    Eye\_L & 1/1   & 3/3   & 2/2   & 6/5   & 5/6   & 4/4   & 9/8   & 8/7   & 7/9   & 10/10 \\
    Eye\_R & 1/1   & 3/3   & 2/4   & 4/2   & 7/7   & 5/5   & 8/8   & 6/9   & 9/6   & 10/10 \\
    Hippocampus\_L & 1/1   & 2/2   & 3/3   & 5/4   & 6/6   & 4/5   & 8/8   & 7/9   & 9/7   & 10/10 \\
    Hippocampus\_R & 1/1   & 3/2   & 5/5   & 4/3   & 2/4   & 7/7   & 8/8   & 6/10  & 10/6  & 9/9 \\
    IAC\_L & 1/1   & 2/2   & 5/5   & 7/7   & 6/6   & 4/4   & 8/8   & 3/10  & 10/3  & 9/9 \\
    IAC\_R & 1/1   & 3/3   & 2/2   & 5/5   & 4/4   & 6/6   & 8/8   & 7/10  & 10/7  & 9/9 \\
    Larynx & 1/1   & 4/3   & 2/2   & 5/5   & 3/4   & 6/6   & 8/8   & 7/9   & 10/7  & 9/10 \\
    Larynx\_Glottic & 1/2   & 2/1   & 3/3   & 5/5   & 4/4   & 6/6   & 8/9   & 7/7   & 10/8  & 9/10 \\
    Larynx\_Supraglot & 1/1   & 2/2   & 4/4   & 5/5   & 3/3   & 6/6   & 8/9   & 7/8   & 9/7   & 10/10 \\
    Lens\_L & 1/1   & 2/3   & 4/2   & 5/4   & 3/6   & 6/5   & 7/7   & 8/10  & 10/8  & 9/9 \\
    Lens\_R & 1/1   & 2/2   & 4/3   & 5/5   & 3/4   & 6/6   & 8/8   & 7/10  & 10/7  & 9/9 \\
    Mandible\_L & 1/1   & 2/2   & 4/4   & 3/3   & 5/5   & 6/6   & 8/8   & 7/7   & 9/9   & 10/10 \\
    Mandible\_R & 1/1   & 2/2   & 7/4   & 4/5   & 3/3   & 5/6   & 8/8   & 6/9   & 10/7  & 9/10 \\
    Mastoid\_L & 2/3   & 1/2   & 3/1   & 4/4   & 5/5   & 6/6   & 7/8   & 8/7   & 9/9   & 10/10 \\
    Mastoid\_R & 1/1   & 2/3   & 6/2   & 4/4   & 3/5   & 5/6   & 8/9   & 7/8   & 10/7  & 9/10 \\
    MiddleEar\_L & 1/1   & 2/2   & 3/3   & 4/4   & 5/5   & 6/6   & 8/7   & 7/9   & 10/8  & 9/10 \\
    MiddleEar\_R & 2/2   & 5/4   & 1/1   & 7/7   & 3/3   & 6/6   & 8/8   & 4/9   & 10/5  & 9/10 \\
    OpticNerve\_L & 1/2   & 3/3   & 7/5   & 4/4   & 5/7   & 8/9   & 2/1   & 9/10  & 10/6  & 6/8 \\
    OpticNerve\_R & 1/1   & 3/3   & 2/2   & 5/4   & 4/5   & 7/7   & 8/8   & 10/6  & 6/10  & 9/9 \\
    OralCavity & 1/1   & 4/5   & 3/3   & 7/7   & 6/4   & 5/6   & 8/8   & 10/2  & 2/10  & 9/9 \\
    Parotid\_L & 4/4   & 2/1   & 3/2   & 7/7   & 6/6   & 5/5   & 8/8   & 10/3  & 1/10  & 9/9 \\
    Parotid\_R & 6/3   & 2/4   & 1/1   & 5/5   & 7/7   & 4/6   & 8/8   & 10/2  & 3/10  & 9/9 \\
    PharynxConst & 3/2   & 2/3   & 1/1   & 5/4   & 8/8   & 4/6   & 7/7   & 10/5  & 6/10  & 9/9 \\
    Pituitary & 3/4   & 2/2   & 1/1   & 4/3   & 7/7   & 6/6   & 8/9   & 10/5  & 5/10  & 9/8 \\
    SpinalCord & 5/4   & 2/2   & 3/3   & 1/1   & 7/6   & 4/5   & 8/8   & 10/7  & 6/10  & 9/9 \\
    Submandibular\_L & 1/1   & 3/3   & 2/2   & 4/4   & 5/5   & 6/6   & 8/7   & 10/8  & 7/10  & 9/9 \\
    Submandibular\_R & 2/2   & 1/1   & 4/4   & 3/3   & 6/6   & 5/5   & 8/8   & 9/7   & 7/9   & 10/10 \\
    TemporalLobe\_L & 2/2   & 4/4   & 1/1   & 3/3   & 6/6   & 5/5   & 7/7   & 8/10  & 10/8  & 9/9 \\
    TemporalLobe\_R & 1/1   & 3/3   & 2/2   & 4/4   & 6/6   & 5/5   & 7/7   & 8/10  & 10/8  & 9/9 \\
    Thyroid & 1/2   & 2/1   & 3/3   & 5/5   & 9/9   & 4/4   & 6/7   & 8/6   & 7/8   & 10/10 \\
    Trachea & 1/1   & 6/6   & 5/5   & 4/4   & 2/2   & 3/3   & 9/8   & 10/7  & 7/10  & 8/9 \\
    TympanicCavity\_L & 1/1   & 3/3   & 2/2   & 5/5   & 6/6   & 4/4   & 7/7   & 8/10  & 10/8  & 9/9 \\
    TMjoint\_L & 2/1   & 3/2   & 5/4   & 6/3   & 4/5   & 8/7   & 7/8   & 9/10  & 10/9  & 1/6 \\
    TMjoint\_R & 1/1   & 4/2   & 2/3   & 5/5   & 7/7   & 6/6   & 3/4   & 8/10  & 10/9  & 9/8 \\
    TympanicCavity\_R & 1/1   & 3/2   & 4/4   & 6/6   & 7/7   & 5/5   & 2/3   & 9/10  & 10/9  & 8/8 \\
    VestibulSemi\_L & 1/1   & 2/3   & 3/2   & 4/4   & 6/5   & 5/6   & 7/7   & 10/8  & 8/10  & 9/9 \\
    VestibulSemi\_R & 3/4   & 1/1   & 4/3   & 2/5   & 9/9   & 5/7   & 6/2   & 7/10  & 10/6  & 8/8 \\
    \midrule
    Overall & 1     & 2     & 3     & 4     & 5     & 6     & 7     & 8     & 9     & 10 \\
    \bottomrule
    \end{tabular}%
    }
  \label{tab:task1_rank}%
\end{table*}%

\begin{figure*}[t]
    \centering
    \includegraphics[width=0.9\textwidth]{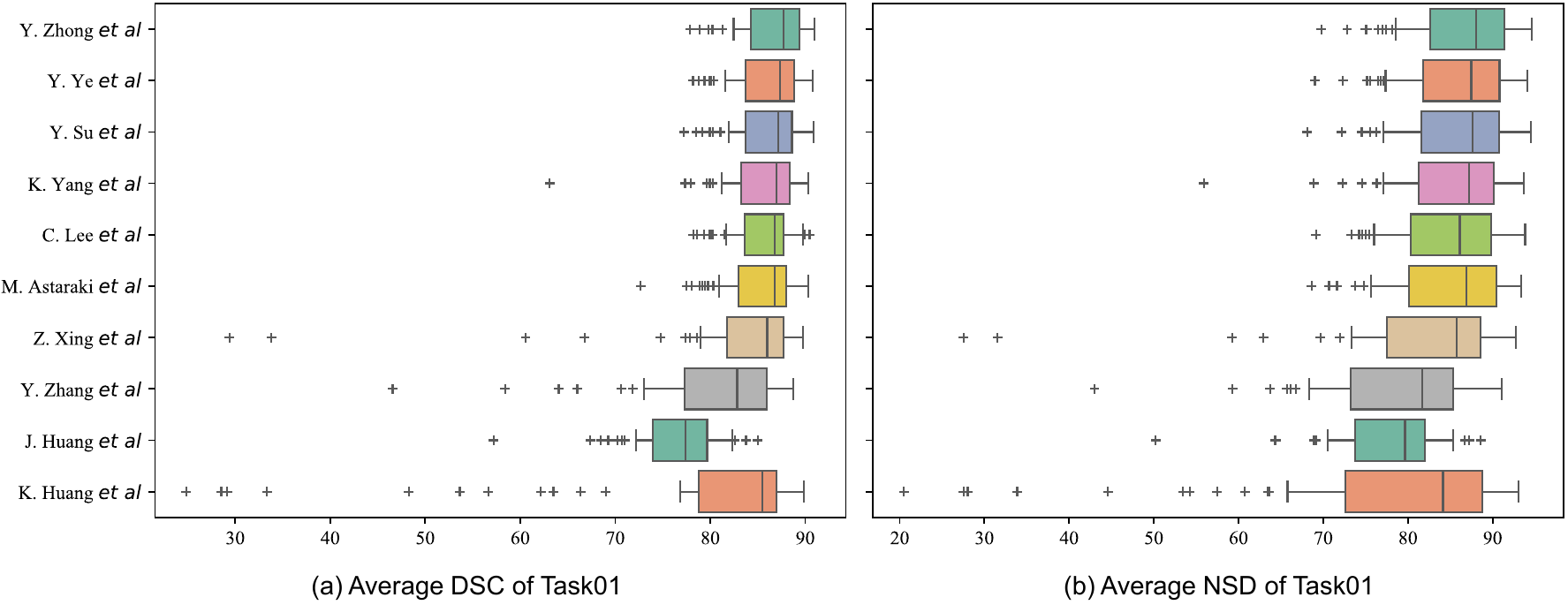}
    \caption{Box plot of the patient-level average segmentation performance for OARs in terms of DSC and NSD.}
    \label{fig:task01_avg_results}
\end{figure*}

\begin{table*}[t]
  \centering
  \caption{Summary of the average DSC score of OAR segmentation by the ten teams.}
  \resizebox{0.9\textwidth}{!}{
    \begin{tabular}{lcccccccccc|cc}
    \toprule
    Team & Y. Zhong \textit{et al.} & Y. Ye \textit{et al.} & Y. Su \textit{et al.} & K. Yang \textit{et al.} & C. Lee \textit{et al.} & M. Astaraki \textit{et al.} & Z. Xing \textit{et al.} & Y. Zhang \textit{et al.} & J. Huang \textit{et al.} & K. Huang \textit{et al.} & Baseline \bigstrut\\
    \hline
    Brain & 98.62±0.26 & 98.63±0.30 & 98.65±0.32 & 98.62±0.31 & 98.58±0.25 & 98.61±0.35 & 98.54±0.22 & 98.44±0.18 & 98.60±0.27 & 98.42±0.22 & 98.47±0.27 \bigstrut[t]\\
    BrainStem & 92.45±2.76 & 92.28±2.67 & 91.97±2.82 & 91.88±2.62 & 91.57±4.45 & 91.75±2.74 & 92.32±2.73 & 92.06±2.77 & 91.92±2.75 & 91.72±2.85 & 91.84±3.01 \\
    Chiasm & 70.55±14.41 & 71.08±13.67 & 69.49±13.34 & 69.67±13.72 & 70.67±15.60 & 70.03±14.41 & 70.53±14.68 & 71.76±13.05 & 64.57±16.07 & 69.13±14.21 & 70.12±12.31 \\
    Cochlea\_L & 94.91±1.36 & 94.77±1.27 & 94.83±1.47 & 94.54±2.13 & 94.76±1.27 & 94.55±1.41 & 87.10±19.12 & 89.02±9.36 & 94.26±1.59 & 83.54±26.02 & 93.27±1.66 \\
    Cochlea\_R & 95.32±1.28 & 94.93±1.53 & 94.99±1.53 & 94.63±2.52 & 94.71±1.42 & 94.84±1.38 & 87.65±18.36 & 88.93±10.50 & 94.52±1.58 & 80.58±30.46 & 94.38±1.73 \\
    Esophagus & 77.32±8.09 & 76.60±7.95 & 77.63±7.81 & 76.69±8.15 & 76.05±8.59 & 75.71±8.10 & 73.53±16.30 & 73.51±9.88 & 73.83±11.55 & 67.91±23.08 & 73.34±9.36 \\
    ETbone\_L & 79.18±8.19 & 78.19±8.20 & 78.98±8.37 & 76.82±12.69 & 77.97±7.91 & 77.38±8.07 & 76.07±16.24 & 77.47±6.59 & 74.55±12.71 & 68.27±26.12 & 77.07±6.88 \\
    ETbone\_R & 94.04±2.09 & 93.91±2.01 & 93.99±2.19 & 93.53±4.74 & 93.69±2.09 & 93.74±2.23 & 88.11±21.67 & 90.00±11.70 & 92.89±4.76 & 84.23±26.48 & 93.14±1.87 \\
    Eye\_L & 93.30±2.08 & 93.17±1.90 & 93.24±2.11 & 91.60±11.29 & 92.72±2.32 & 92.82±2.07 & 87.92±20.51 & 89.23±10.42 & 90.71±12.00 & 81.23±29.61 & 92.52±2.02 \\
    Eye\_R & 78.34±7.78 & 78.02±8.12 & 78.18±8.21 & 77.72±8.99 & 74.78±12.34 & 77.41±8.28 & 73.43±20.77 & 75.10±13.17 & 70.34±15.76 & 67.93±22.86 & 75.4±10.38 \\
    Hippocampus\_L & 75.83±8.52 & 75.54±7.88 & 75.31±7.30 & 74.88±12.74 & 73.31±10.89 & 75.02±7.95 & 71.74±18.55 & 71.95±14.31 & 67.18±18.18 & 64.19±24.61 & 75.29±6.91 \\
    Hippocampus\_R & 79.99±7.71 & 78.99±8.05 & 78.43±8.86 & 78.60±9.48 & 79.44±7.34 & 77.48±9.19 & 75.73±18.94 & 77.75±12.85 & 65.90±20.71 & 69.79±25.48 & 78.49±8.13 \\
    IAC\_L & 81.94±7.23 & 81.75±7.50 & 80.50±8.92 & 79.26±13.27 & 80.24±7.85 & 80.57±7.43 & 78.18±16.92 & 81.01±7.93 & 65.89±24.97 & 71.09±25.13 & 78.59±8.60 \\
    IAC\_R & 88.42±5.18 & 87.38±5.32 & 87.45±4.72 & 86.78±5.93 & 87.16±5.12 & 85.25±7.49 & 82.46±17.02 & 82.68±17.50 & 69.85±7.35 & 76.40±24.44 & 84.85±5.09 \\
    Larynx & 89.25±5.02 & 87.37±5.28 & 87.98±5.08 & 86.62±7.24 & 87.47±6.55 & 85.98±7.74 & 83.10±16.66 & 84.07±15.80 & 68.19±8.82 & 74.56±30.97 & 87.26±4.35 \\
    Larynx\_Glottic & 84.94±8.45 & 84.54±8.13 & 83.82±8.01 & 82.80±9.32 & 83.70±7.62 & 82.36±8.66 & 74.23±17.56 & 79.66±17.29 & 72.73±18.33 & 73.46±23.09 & 94.63±6.46 \\
    Larynx\_Supraglot & 85.34±7.34 & 84.72±7.27 & 84.17±7.33 & 82.28±13.03 & 84.60±6.18 & 81.25±8.88 & 75.82±17.41 & 79.70±19.63 & 70.28±23.08 & 67.76±31.81 & 82.58±8.15 \\
    Lens\_L & 81.95±7.28 & 81.39±7.41 & 80.77±8.17 & 80.64±7.51 & 81.00±7.30 & 80.27±8.49 & 76.96±16.47 & 74.80±20.48 & 52.98±11.99 & 71.39±23.57 & 78.62±9.20 \\
    Lens\_R & 84.18±7.22 & 83.58±7.15 & 82.83±7.63 & 82.33±7.76 & 83.57±7.16 & 81.57±8.06 & 78.96±16.66 & 79.39±16.33 & 55.07±13.34 & 70.78±28.94 & 82.47±7.64 \\
    Mandible\_L & 83.79±8.80 & 83.42±8.51 & 82.68±8.47 & 82.75±9.03 & 82.38±7.77 & 81.67±11.81 & 77.55±17.46 & 77.98±20.02 & 73.33±14.70 & 71.63±24.32 & 82.39±8.03 \\
    Mandible\_R & 83.49±9.06 & 83.19±8.55 & 79.35±10.92 & 82.25±9.04 & 82.65±7.60 & 81.07±12.63 & 77.98±15.78 & 79.48±16.41 & 66.84±18.83 & 67.28±29.47 & 82.49±8.14 \\
    Mastoid\_L & 84.10±8.21 & 84.50±7.72 & 84.04±7.42 & 83.49±8.23 & 82.56±8.01 & 81.81±12.57 & 78.98±16.85 & 78.25±20.06 & 72.57±18.13 & 71.46±24.56 & 82.92±8.47 \\
    Mastoid\_R & 83.35±9.43 & 82.85±9.47 & 80.43±11.63 & 81.50±13.63 & 81.97±8.31 & 80.98±12.45 & 76.76±16.70 & 79.54±16.75 & 68.09±22.35 & 68.15±29.63 & 82.52±9.48 \\
    MiddleEar\_L & 82.14±5.72 & 82.06±5.49 & 81.46±5.72 & 80.92±6.77 & 80.46±7.23 & 79.77±7.90 & 77.36±15.87 & 77.64±17.39 & 66.98±16.86 & 72.24±23.18 & 70.65±8.31 \\
    MiddleEar\_R & 78.99±10.86 & 76.35±9.74 & 79.12±9.46 & 74.61±12.70 & 78.06±9.95 & 74.78±10.87 & 74.40±16.81 & 76.54±14.28 & 61.84±18.19 & 67.83±25.41 & 74.82±9.83 \\
    OpticNerve\_L & 77.70±13.86 & 77.27±13.6 & 75.78±17.65 & 76.58±16.14 & 76.58±16.31 & 75.52±14.98 & 77.65±14.04 & 75.35±17.87 & 64.44±23.53 & 75.78±13.26 & 75.81±16.44 \\
    OpticNerve\_R & 95.04±1.56 & 94.96±1.61 & 94.98±1.64 & 94.94±1.60 & 94.95±1.59 & 94.79±1.58 & 94.63±1.61 & 94.15±1.70 & 94.85±1.57 & 94.28±1.79 & 93.89±1.78 \\
    OralCavity & 95.02±1.88 & 94.92±1.84 & 94.99±1.87 & 92.60±3.74 & 94.35±2.04 & 94.67±1.89 & 90.47±15.26 & 72.19±19.30 & 95.01±1.90 & 85.46±22.03 & 93.38±2.30 \\
    Parotid\_L & 94.27±3.30 & 94.39±3.23 & 94.36±3.33 & 91.73±6.08 & 93.76±3.32 & 94.16±3.22 & 91.10±12.85 & 73.17±18.57 & 94.41±3.15 & 84.57±22.07 & 93.41±3.41 \\
    Parotid\_R & 88.99±9.85 & 89.63±6.48 & 89.74±6.26 & 89.00±7.61 & 87.94±9.33 & 89.13±7.73 & 86.70±13.62 & 67.10±20.60 & 89.30±7.19 & 83.82±18.29 & 88.31±7.54 \\
    PharynxConst & 87.27±11.50 & 87.59±9.24 & 87.82±9.18 & 86.46±12.04 & 85.11±13.41 & 87.23±9.48 & 85.65±13.63 & 66.49±21.86 & 85.94±15.49 & 81.91±18.79 & 86.99±9.12 \\
    Pituitary & 90.26±4.41 & 90.28±4.51 & 90.36±4.66 & 90.25±4.48 & 89.09±5.32 & 89.89±4.52 & 83.04±16.22 & 70.21±24.19 & 90.23±4.49 & 81.62±24.54 & 88.36±5.28 \\
    SpinalCord & 88.26±7.46 & 88.68±6.50 & 88.63±6.33 & 88.99±5.73 & 86.41±11.02 & 88.46±6.46 & 82.39±16.29 & 71.96±20.22 & 87.40±7.22 & 78.44±24.69 & 86.32±7.56 \\
    Submandibular\_L & 92.90±2.40 & 92.79±2.58 & 92.84±2.53 & 92.55±2.70 & 92.36±2.50 & 92.33±2.67 & 84.56±19.04 & 79.26±23.89 & 86.69±4.59 & 81.31±26.09 & 90.62±3.92 \\
    Submandibular\_R & 92.47±3.52 & 92.49±3.49 & 92.30±3.40 & 92.35±3.60 & 92.00±3.63 & 92.05±3.64 & 84.46±19.79 & 82.66±16.45 & 87.95±4.30 & 77.68±30.37 & 91.62±3.69 \\
    TemporalLobe\_L & 89.23±7.20 & 88.84±7.08 & 89.32±6.80 & 88.88±7.36 & 88.45±7.40 & 88.54±7.06 & 81.76±21.86 & 79.91±23.90 & 73.35±19.82 & 79.19±25.61 & 88.37±6.81 \\
    TemporalLobe\_R & 90.37±4.72 & 89.72±5.17 & 89.95±4.69 & 89.43±5.55 & 88.78±6.09 & 89.21±5.89 & 83.88±15.17 & 83.32±15.93 & 67.09±22.58 & 75.22±31.10 & 89.22±4.53 \\
    Thyroid & 89.69±4.29 & 89.54±3.85 & 89.44±3.98 & 89.27±4.05 & 88.80±4.14 & 89.28±4.12 & 89.17±4.21 & 88.90±3.96 & 88.95±3.66 & 88.32±4.00 & 88.52±3.31 \\
    TMjoint\_L & 82.34±8.16 & 82.25±8.01 & 82.21±8.00 & 81.86±8.01 & 82.21±8.00 & 81.31±8.51 & 81.41±7.98 & 81.26±7.56 & 34.91±25.87 & 82.42±7.97 & 84.33±10.96 \\
    TMjoint\_R & 89.74±3.97 & 89.28±4.18 & 89.35±3.91 & 89.14±4.19 & 88.75±3.89 & 88.90±3.98 & 89.32±3.95 & 88.35±3.95 & 63.13±23.27 & 88.08±3.45 & 89.59±4.41 \\
    Trachea & 85.01±2.66 & 83.98±2.15 & 84.07±2.26 & 84.08±2.20 & 84.81±2.91 & 84.10±2.09 & 82.57±3.50 & 82.28±3.11 & 82.89±3.47 & 82.70±2.94 & 79.65±4.65 \\
    TympanicCavity\_L & 89.66±2.21 & 89.37±2.18 & 89.55±2.32 & 89.23±2.38 & 89.21±2.07 & 89.25±2.43 & 89.03±2.37 & 88.80±2.17 & 81.43±4.94 & 88.76±2.20 & 88.43±2.03 \\
    TympanicCavity\_R & 85.17±4.83 & 84.53±4.66 & 84.36±4.89 & 84.04±4.92 & 83.03±4.57 & 84.08±4.85 & 84.71±4.89 & 81.05±6.05 & 71.91±6.89 & 82.13±5.85 & 81.77±3.83 \\
    VestibulSemi\_L & 91.27±3.34 & 90.90±3.11 & 90.90±3.15 & 90.59±3.12 & 90.25±3.44 & 90.30±3.07 & 90.10±3.16 & 88.66±3.83 & 89.91±3.36 & 88.84±3.19 & 79.46±9.08 \\
    VestibulSemi\_R & 85.18±9.46 & 85.56±8.55 & 85.11±8.96 & 85.48±7.87 & 84.46±9.47 & 84.96±8.85 & 84.94±9.22 & 84.73±8.50 & 77.13±7.37 & 84.69±8.46 & 84.27±6.97 \bigstrut[b]\\
    \hline
    Average & 86.70±9.30 & 86.36±9.15 & 86.14±9.58 & 85.62±10.48 & 85.68±9.87 & 85.44±10.17 & 82.51±16.48 & 80.57±16.52 & 76.68±19.62 & 78.14±23.65 & 84.65±9.95 \bigstrut\\
    \hline
    \end{tabular}%
    }
  \label{tab:task01_DSC}%
\end{table*}%

\begin{table*}[t]
  \centering
  \caption{Summary of the average NSD score of OAR segmentation by the ten teams.}
    \resizebox{0.9\textwidth}{!}{
    \begin{tabular}{lcccccccccc|cc}
    \toprule
    Team & Y. Zhong \textit{et al.} & Y. Ye \textit{et al.} & Y. Su \textit{et al.} & K. Yang \textit{et al.} & C. Lee \textit{et al.} & M. Astaraki \textit{et al.} & Z. Xing \textit{et al.} & Y. Zhang \textit{et al.} & J. Huang \textit{et al.} & K. Huang \textit{et al.} & Baseline \bigstrut\\
    \hline
    Brain & 89.68±4.75 & 89.77±5.25 & 89.79±5.28 & 89.64±5.29 & 88.92±4.84 & 89.39±5.81 & 88.92±4.80 & 87.57±4.56 & 88.89±5.11 & 87.08±5.06 & 88.02±4.93 \bigstrut[t]\\
    BrainStem & 82.00±10.57 & 81.54±10.29 & 80.57±10.65 & 80.27±9.95 & 79.28±12.01 & 79.82±10.12 & 81.55±10.60 & 80.38±10.28 & 80.17±10.40 & 79.16±11.02 & 79.13±11.03 \\
    Chiasm & 77.07±15.58 & 77.50±14.65 & 75.98±14.4 & 76.18±14.36 & 75.84±17.35 & 76.38±15.59 & 76.77±16.05 & 77.24±13.85 & 72.55±15.29 & 75.18±14.71 & 75.79±12.76 \\
    Cochlea\_L & 79.99±7.71 & 79.43±7.04 & 79.76±7.43 & 78.90±7.67 & 79.26±7.24 & 78.31±7.73 & 69.98±16.97 & 70.37±10.53 & 77.11±7.19 & 66.35±22.97 & 73.14±8.17 \\
    Cochlea\_R & 80.61±7.80 & 78.60±8.85 & 78.91±8.78 & 78.21±9.29 & 77.57±8.87 & 77.99±8.46 & 69.33±16.95 & 68.59±12.74 & 76.28±8.81 & 63.14±25.63 & 76.50±8.57 \\
    Esophagus & 68.31±12.25 & 67.57±12.03 & 68.92±11.76 & 68.06±11.69 & 66.24±12.48 & 66.14±11.74 & 64.93±16.76 & 62.79±13.27 & 64.89±13.97 & 59.43±21.29 & 60.88±12.99 \\
    ETbone\_L & 71.62±13.33 & 70.06±12.92 & 71.31±13.82 & 68.81±15.40 & 68.88±13.05 & 68.69±12.98 & 68.02±18.13 & 68.48±11.05 & 65.67±14.79 & 61.14±24.81 & 68.08±11.17 \\
    ETbone\_R & 91.16±8.21 & 90.87±7.64 & 90.81±8.41 & 90.73±9.33 & 89.84±8.07 & 90.01±8.49 & 84.88±21.83 & 85.43±14.38 & 88.78±9.28 & 79.67±26.71 & 88.49±7.27 \\
    Eye\_L & 88.71±8.11 & 88.40±7.77 & 88.66±9.12 & 87.15±11.96 & 86.89±8.99 & 87.43±8.16 & 83.00±20.97 & 82.79±13.17 & 83.94±13.79 & 74.40±28.73 & 86.37±8.03 \\
    Eye\_R & 90.12±8.76 & 89.88±8.71 & 89.64±8.66 & 89.90±9.64 & 85.80±12.36 & 89.07±8.74 & 84.35±22.90 & 86.61±13.24 & 82.47±17.22 & 76.96±26.45 & 87.61±10.22 \\
    Hippocampus\_L & 86.58±10.94 & 86.42±10.33 & 86.38±8.96 & 85.63±15.20 & 83.46±12.23 & 85.53±10.29 & 81.77±21.68 & 83.15±13.58 & 78.56±20.00 & 71.97±28.12 & 86.11±8.85 \\
    Hippocampus\_R & 87.96±9.00 & 86.73±9.32 & 86.20±10.42 & 86.41±10.69 & 86.30±8.23 & 84.95±10.63 & 83.16±20.83 & 85.39±14.38 & 76.88±17.87 & 77.39±27.81 & 85.54±9.59 \\
    IAC\_L & 89.19±7.66 & 89.16±7.88 & 87.84±9.28 & 86.74±14.40 & 86.88±8.54 & 87.86±8.16 & 84.90±18.33 & 88.01±8.65 & 75.94±23.33 & 78.10±26.91 & 84.83±9.74 \\
    IAC\_R & 91.71±6.54 & 90.28±6.99 & 90.43±5.98 & 89.68±7.50 & 89.94±6.61 & 87.84±9.26 & 85.11±17.75 & 85.59±17.44 & 75.54±5.58 & 78.33±25.03 & 87.46±6.33 \\
    Larynx & 98.10±3.54 & 97.03±3.92 & 97.53±2.84 & 96.54±5.56 & 96.63±5.00 & 96.09±6.84 & 92.2±17.52 & 93.75±14.35 & 86.65±4.65 & 85.11±30.30 & 83.5±8.22 \\
    Larynx\_Glottic & 95.38±6.24 & 95.40±6.06 & 95.07±6.26 & 94.19±8.04 & 94.43±6.41 & 93.72±7.39 & 86.18±18.62 & 90.37±18.57 & 90.94±13.52 & 84.29±23.52 & 97.38±2.48 \\
    Larynx\_Supraglot & 96.17±5.44 & 95.88±5.85 & 95.67±5.53 & 93.90±13.54 & 95.86±4.75 & 93.48±7.57 & 87.69±18.65 & 90.73±19.27 & 87.72±20.38 & 78.02±35.78 & 94.59±6.69 \\
    Lens\_L & 92.05±6.83 & 91.69±7.02 & 91.71±6.91 & 91.17±6.82 & 90.42±6.95 & 90.57±8.62 & 86.53±17.63 & 85.38±20.46 & 67.51±10.57 & 80.40±26.61 & 88.52±7.95 \\
    Lens\_R & 92.27±7.17 & 91.66±7.46 & 91.31±7.94 & 91.10±8.19 & 91.18±8.32 & 90.24±8.91 & 86.63±18.03 & 87.47±15.81 & 67.26±12.83 & 78.62±29.98 & 90.19±7.99 \\
    Mandible\_L & 94.99±7.19 & 94.97±6.98 & 94.72±7.16 & 94.83±7.22 & 93.93±7.46 & 93.66±9.64 & 88.82±17.91 & 88.80±21.98 & 89.49±10.39 & 83.58±25.60 & 94.55±6.85 \\
    Mandible\_R & 94.94±6.94 & 94.85±6.74 & 94.58±6.38 & 94.23±7.32 & 94.65±6.60 & 93.68±9.61 & 90.06±16.37 & 91.26±15.45 & 84.97±14.03 & 78.89±32.05 & 94.84±6.58 \\
    Mastoid\_L & 95.43±6.70 & 95.69±6.14 & 95.84±5.79 & 95.11±7.39 & 93.97±6.63 & 93.90±10.61 & 90.75±18.45 & 88.93±21.90 & 92.12±10.52 & 83.16±27.24 & 94.35±6.83 \\
    Mastoid\_R & 95.10±7.91 & 94.70±8.12 & 94.88±7.00 & 93.99±11.00 & 93.84±7.46 & 93.19±10.38 & 89.04±17.92 & 91.50±15.78 & 89.27±14.63 & 80.41±30.46 & 94.53±8.08 \\
    MiddleEar\_L & 95.01±4.37 & 94.93±4.47 & 94.87±4.52 & 94.19±5.78 & 93.93±6.02 & 93.39±6.28 & 90.40±17.2 & 90.33±18.99 & 87.38±13.56 & 84.86±25.40 & 86.70±6.76 \\
    MiddleEar\_R & 92.60±9.06 & 91.16±7.60 & 93.41±7.43 & 89.55±11.08 & 91.37±9.40 & 89.85±9.24 & 88.44±17.09 & 90.55±13.75 & 84.19±14.82 & 81.35±27.67 & 89.93±7.94 \\
    OpticNerve\_L & 86.61±11.99 & 85.90±11.99 & 84.73±16.59 & 84.78±16.08 & 84.56±15.9 & 84.34±13.38 & 86.69±12.24 & 84.69±14.77 & 79.65±21.20 & 84.43±11.96 & 84.59±15.79 \\
    OpticNerve\_R & 75.79±10.36 & 75.32±10.46 & 75.48±10.55 & 75.09±10.34 & 74.85±10.81 & 74.26±9.98 & 72.94±10.54 & 69.66±10.86 & 74.84±10.29 & 70.94±11.69 & 69.01±10.45 \\
    OralCavity & 99.79±0.56 & 99.72±0.47 & 99.74±0.47 & 96.88±3.37 & 99.74±0.60 & 99.72±0.52 & 95.43±15.02 & 76.82±13.87 & 99.75±0.39 & 91.04±21.56 & 98.77±1.48 \\
    Parotid\_L & 91.87±9.84 & 92.17±9.33 & 92.07±9.67 & 88.57±10.41 & 89.93±11.59 & 91.72±9.53 & 87.82±15.78 & 65.97±17.64 & 92.02±9.50 & 80.89±23.70 & 89.58±9.51 \\
    Parotid\_R & 82.91±14.67 & 82.79±12.91 & 83.53±12.02 & 82.42±13.51 & 79.86±13.89 & 82.00±13.95 & 79.70±17.06 & 58.44±17.61 & 83.40±12.46 & 75.35±19.66 & 78.75±15.08 \\
    PharynxConst & 79.34±18.10 & 79.08±16.74 & 79.66±16.66 & 78.32±18.06 & 74.14±18.84 & 77.87±17.67 & 76.52±18.75 & 55.21±20.20 & 78.30±19.54 & 72.43±21.29 & 76.45±16.94 \\
    Pituitary & 74.12±15.93 & 74.14±16.30 & 74.51±16.71 & 74.12±16.19 & 70.20±16.50 & 72.50±16.29 & 65.99±19.65 & 56.07±19.93 & 73.74±16.30 & 66.91±24.45 & 68.53±16.03 \\
    SpinalCord & 70.25±19.22 & 71.21±18.31 & 70.84±18.64 & 71.53±17.16 & 68.13±18.61 & 69.95±18.42 & 62.98±20.18 & 53.59±18.92 & 66.39±20.17 & 60.69±24.24 & 64.62±18.47 \\
    Submandibular\_L & 90.06±6.36 & 89.68±7.09 & 89.86±6.94 & 89.25±7.04 & 88.87±6.23 & 88.65±7.18 & 79.47±18.82 & 75.79±22.20 & 79.26±8.48 & 76.41±26.28 & 84.18±8.71 \\
    Submandibular\_R & 88.93±9.12 & 89.13±8.97 & 88.68±8.89 & 88.69±9.14 & 87.71±9.61 & 87.96±9.33 & 78.94±20.22 & 77.12±18.07 & 80.10±8.82 & 73.53±28.56 & 86.79±9.11 \\
    TemporalLobe\_L & 87.63±12.37 & 87.18±12.19 & 87.95±11.72 & 87.27±12.41 & 86.22±12.68 & 86.69±12.00 & 79.67±22.85 & 78.71±23.43 & 74.70±17.31 & 77.11±26.57 & 86.27±11.54 \\
    TemporalLobe\_R & 89.89±8.22 & 88.93±9.00 & 89.26±8.24 & 88.38±9.64 & 86.60±11.24 & 87.81±10.27 & 81.53±17.32 & 81.52±17.72 & 72.94±17.02 & 74.25±29.92 & 87.79±8.14 \\
    Thyroid & 86.53±11.01 & 86.72±10.41 & 86.39±10.92 & 86.09±10.95 & 84.34±10.97 & 86.13±11.00 & 85.36±11.15 & 84.96±10.89 & 86.06±10.09 & 83.76±11.24 & 84.62±10.06 \\
    TMjoint\_L & 80.14±12.74 & 80.05±12.56 & 79.65±12.71 & 79.67±12.52 & 79.51±12.85 & 78.90±12.90 & 78.45±13.18 & 77.54±11.69 & 35.72±23.12 & 78.98±12.27 & 77.97±14.18 \\
    TMjoint\_R & 88.36±7.82 & 87.89±7.74 & 87.88±7.31 & 87.33±7.75 & 86.88±7.63 & 87.14±7.45 & 87.54±7.66 & 85.69±7.51 & 60.06±21.32 & 86.12±7.07 & 86.81±6.75 \\
    Trachea & 78.04±5.72 & 75.18±5.83 & 75.29±5.99 & 75.30±5.88 & 77.00±6.11 & 75.43±6.20 & 72.45±6.98 & 71.51±6.74 & 73.97±8.95 & 71.76±7.44 & 68.10±7.98 \\
    TympanicCavity\_L & 75.71±9.08 & 74.86±8.49 & 75.12±9.40 & 73.72±9.52 & 72.54±8.73 & 74.25±8.92 & 72.38±9.43 & 71.31±8.80 & 60.59±8.78 & 71.30±9.25 & 69.36±7.68 \\
    TympanicCavity\_R & 86.41±9.70 & 85.86±9.24 & 85.70±9.89 & 84.77±9.92 & 81.92±9.81 & 85.15±9.46 & 85.76±9.97 & 79.19±12.11 & 72.22±10.29 & 81.37±12.48 & 80.91±9.54 \\
    VestibulSemi\_L & 89.19±9.27 & 88.36±8.94 & 88.58±9.16 & 87.78±9.05 & 87.08±9.42 & 86.94±9.34 & 86.50±9.37 & 83.03±10.55 & 86.02±9.27 & 83.27±10.44 & 67.15±12.31 \\
    VestibulSemi\_R & 75.36±17.45 & 75.87±15.43 & 75.51±16.01 & 74.97±14.70 & 72.40±18.46 & 74.68±16.49 & 75.72±16.48 & 74.74±15.35 & 58.30±12.54 & 74.66±15.13 & 71.12±14.15 \bigstrut[b] \\
    \hline
    Average & 86.53±12.85 & 86.09±12.64 & 86.12±12.79 & 85.33±13.42 & 84.62±13.62 & 84.96±13.21 & 81.67±18.56 & 79.18±18.69 & 77.85±18.04 & 76.94±24.31 & 82.88±14.01 \bigstrut\\
    \hline
    \end{tabular}%
    }
  \label{tab:task01_NSD}%
\end{table*}%

\subsection{Task01: OAR segmentation}
\begin{itemize}[label=\textcolor{red}{\textbullet}]
  \item ($1^{st}$ place, Y. Zhong \textit{et al.}) Zhong \textit{et al.} proposed a two-step approach to segment OARs: structure-specific label generation and boundary refinement. For structure-specific label generation, 45 organs are divided into 29 distinct classes considering the left and right counterparts and label overlapping in the ear and oral cavity. The segmentation model was built based on nnUNetV2~\citep{isensee2021nnu} and trained with paired non-contrast and contrast-enhancement CT scans. For the boundary refinement, ROI with a size of $128 \times 128 \times 128$ was extracted based on the segmentation result. The refinement model has an encoder, a decoder, and multiple output layers. All organs share the same encoder and decoder weights within the model but employ different output layers. Finally, the refined ROI was then integrated back into the original segmentation results. 

  \item ($2^{nd}$ place, Y. Ye \textit{et al.}) This method was based on UniSeg~\citep{Ye2023UniSeg} and bespoke pre-processing and ensemble strategy. The UniSeg model is a supervised pre-trained nnUNet model, which is trained on 11 3D partially labeled segmentation datasets spanning multiple targets, domains, and modalities. To fine-tune the UniSeg model to OAR segmentation, the images were first pre-processed following nnUNet~\citep{isensee2021nnu} and then resampled to match the median spacing. Then, UniSeg was trained with 1500 epochs and 2000 epochs using paired ncCT and ceCT images. At the inference stage, the given image was pre-processed with nnUNet’s pre-processing step, then segmented into patches using a sliding window approach, and the two predictions for each patch from two fine-tuned UniSeg models were averaged to form the final segmentation map.
  
  \item ($3^{rd}$ place, Y. Su \textit{et al.}) This method used vanilla nnUNet~\citep{isensee2021nnu} to perform OAR segmentation. Data augmentation techniques were used during training, including additive brightness, gamma, rotation, scaling, and elastic deformation. Mirror operation was not used because of the high symmetry of organs in the head and neck. To improve the model's performance, an increased patch size ($48 \times 256 \times 256$) was used during training.
  
  \item ($4^{th}$ place, K. Yang \textit{et al.}) This method used nnUNet~\citep{isensee2021nnu} and region-based training mode for accurate and efficient segmentation. In the training stage, mirror data augmentation was not used, but elastic deformation instead. Masked loss function was used to solve the label missing problem, where the channels of label missing were ignored to correct model training. To solve the overlapping problem, the region-based training mode was used to segment areas that are merged by more than one class. In the inference stage, sliding window strategy and a connect component-based post-processing algorithm were adopted to obtain segmentation results of whole CT images. The code is available at: \url{https://github.com/Kaixiang-Yang/SegRap23}. 
  
  \item ($5^{th}$ place, C. Lee \textit{et al.}) Lee \textit{et al.} proposed a two-step method: localization and segmentation. In the localization stage, a bounding box was identified to encompass the OARs utilizing a 2D-based object detection network powered by the YOLO-v7 model~\citep{wang2022efficient}. In the segmentation stage, different window widths and levels were used for multi-channel input generation. Single organ training and symmetrical OARs Flipped-Unification were used to train segmentation networks with DynUNet architecture using these multi-channel inputs. For OARs Flipped-Unification, the training data was from one of the symmetrical OARs while utilizing a flipped version of the same to represent its counterpart because of the symmetry in the head and neck area. Finally, five models were trained with different patch sizes, localization training, and symmetrical OARs unification. In the inference stage, ROIs were first extracted, and then all predictions from five segmentation models were averaged as final results. 
  
  \item ($6^{th}$ place, M. Astaraki \textit{et al.}) This method was based on harmonizing the intensity distribution and efficient cropping. To better distinguish the overlapping OARs from each other, the HU values of the ceCT and ncCT volumes were clamped into the range of [-400, 2000] and [-300, 800] for pre-processing, respectively. The pre-processed paired full-resolution CT images were used to train a segmentation network based on the nnUNetV1~\citep{isensee2021nnu} framework with 2000 epochs using five-fold cross-validation fashion. In the inference stage, the volumes were cropped based on the TotalSegmentor~\citep{Wasserthal2023TotalSegmentator} model and a connected component analysis and sent to the segmentation network for segmentation labels over the cropped images. The code is available at: \url{https://github.com/Astarakee/segrap2023}.
  
  \item ($7^{th}$ place, Z. Xing \textit{et al.}) This method used crop and test time augmentation strategies to perform OAR segmentation. To save training time, the pixels whose intensity value is out of [-175, 250] were filtered out. Extensive data augmentation operations, including spatial and intensity transforms, were used to improve the robustness segmentation model. Five segmentation models based on UNet structure with different batch sizes, parameter scales, and normalization methods were used to generate a robust prediction. In the test phase, test time augmentation, such as mirror operation and overlapped windows inference, was used to improve the robustness of the prediction. 
  
  \item ($8^{th}$ place, Y. Zhang \textit{et al.}) This method was based on nnUNet~\citep{isensee2021nnu} framework. The HU values of the CT images were clipped to [0.5, 99.5]. Data augmentation methods, including spatial-, intensity- and label-based transformation, were used to enhance data diversity and richness. Paired CT images were randomly cropped into patches of size [28, 224, 224] and used to train a 3D full-resolution UNet based on nnUNet~\citep{isensee2021nnu}. In the inference stage, the patch size was equal to the patch size during training, and the neighbourhood inference was performed using intervals of 1/2 patch size. 
  
  \item ($9^{th}$ place, J. Huang \textit{et al.}) This method used two progressive steps for OAR segmentation: coarse segmentation and fine segmentation. The values of paired CT images were clipped to [-300, 1500] and then normalized to [-1, 1] by min-max normalization. Data augmentation methods like random flipping and rotation were used. In the coarse segmentation stage, pre-processed images were used to train a 3D UNet to get the position and size of the target areas. Then, the corresponding ROIs were cropped based on the coarse segmentation results. In the fine stage, a 3D UNet was trained based on paired CT images and corresponding ROIs to make fine adjustments to the coarse segmentation results.

  \item ($10^{th}$ place, K. Huang \textit{et al.}) This method was based on the nnUNetV2 framework~\citep{isensee2021nnu}. The paired CT volumes were resampled, cropped, and normalized following~\citep{isensee2021nnu}. Data augmentation strategies, including spatial transform, intensity transform, simulate low-resolution transform, were used to improve the diversity of data. Five-fold cross-validation was used to train segmentation networks. In the inference stage, various augmentations like different region cropping and adjustments in scaling were applied, and the average of predictions was taken as the final results.  
\end{itemize}

\subsection{Task02: GTV segmentation}
\begin{itemize}[label=\textcolor{blue}{\textbullet}]
  \item ($1^{st}$ place, M. Astaraki \textit{et al.}) This method was based on intensity distribution harmonization and efficient cropping. In the pre-processing stage, the HU values of the ceCT and ncCT volumes were clamped into the range of [-1000, 1000] and [-600, 600], respectively, to better distinguish the pathological regions from nearby healthy tissues. To discard the background and irrelevant anatomical structures, the paired CT volumes were cropped based on TotalSegmentor~\citep{Wasserthal2023TotalSegmentator} model and a connected component analysis. The cropped paired CT images were used to train a segmentation network based on the nnUNetV1~\citep{isensee2021nnu} framework with 600 epochs using five-fold cross-validation fashion. In the inference stage, the volumes were pre-processed as training data and then sent to the segmentation network for segmentation labels over the cropped images. 

  \item ($2^{nd}$ place, Y. Ye \textit{et al.}) This method was based on UniSeg~\citep{Ye2023UniSeg} and bespoke pre-processing and ensemble strategy. In the training stage, each image was divided into multiple 3D patches of identical size using a sliding window approach, and then these patches were pre-processed following nnUNet~\citep{isensee2021nnu}. Then, UniSeg was trained using paired patches with 1000 epochs. In the inference stage, the entire image was segmented into multiple overlapping patches, and then each patch was sent to the fine-tuned UniSeg to predict its corresponding segmentation map, and these individual patch-based predictions were aggregated as the final prediction.
  
  \item ($3^{rd}$ place, Z. Xing \textit{et al.}) This method used crop and test time augmentation strategies to perform GTV segmentation. Useless areas were cropped based on the HU values. To improve the robustness of the segmentation model, spatial- and intensity-based transforms are used. Five segmentation models based on UNet structure with different batch sizes, parameter scales, and normalization methods were used to generate a robust prediction. In the test phase, test time augmentation was used to improve the robustness of the prediction.
  
  \item ($4^{th}$ place, K. Yang \textit{et al.}) This method was based on nnUNet~\citep{isensee2021nnu} to perform GTV segmentation. Owing to the large variances among patients, Dice loss and Focal loss~\citep{lin2017focal} were used to make the segmentation model focus on the GTVs that are not prone to distinguish. In the inference stage, flipping test time augmentation was used to improve the segmentation performance. The code is available at: \url{https://github.com/Kaixiang-Yang/SegRap23}. 
  
  \item ($5^{th}$ place, C. Ulrich \textit{et al.}) This method employed MultiTalent~\citep{Ulrich2023MultiTalent} model that is trained with multiple partially labeled datasets. Based on the target spacing, normalization scheme, and network topology suggested by nnUNet experiment planning for the SegRap2023, the MultiTalent model was pre-trained with multiple partially annotated datasets. Then the pre-trained MultiTalent model was fine-tuned with paired CT images by only updating the segmentation heads for 10 epochs, and the whole network was updated for 50 epoch warm-up period. Finally, a Residual Encoder (Resenc) UNet was initialized using the MultiTalent model and trained with 2000 epochs to generate the final segmentation results. 
  
  \item ($6^{th}$ place, N. Ndipenoch \textit{et al.}) N. Ndipenoch \textit{et al.} proposed a nnUNet with squeeze and excitation block (nnUNet\_SE) model. There are two modifications in nnUNet\_SE~\citep{isensee2021nnu}: residual blocks are introduced to mitigate the problem of vanishing gradients, and the squeeze-and-excitation block is introduced to capture global features. The nnUNet\_SE model was trained with paired ncCT and ceCT scans, and each of the GTVs was trained separately (i.e., binary segmentation) to improve the performance.

  \item ($7^{th}$ place, Y. Su \textit{et al.}) This method used vanilla nnUNet~\citep{isensee2021nnu} to perform GTV segmentation. Almost all settings were the same as those automatically generated following~\citep{isensee2021nnu}, except for the patch size. A large patch size ($48 \times 256 \times 256$) was used to improve the model’s performance. In the inference stage, test time augmentation strategy was applied for robust segmentation results.
  
  \item ($8^{th}$ place, J. Huang \textit{et al.}) This method used two progressive steps for GTV segmentation: coarse segmentation and fine segmentation. The HU values of paired CT images were clipped to [-300, 1500] and then normalized to [-1, 1] by min-max normalization. Data augmentation methods like random flip and rotate were used. In the coarse segmentation stage, the recall rate was improved as much as possible to segment tumor areas. Then, the corresponding tumor regions were cropped based on the coarse segmentation results. In the fine stage, a 3D UNet was trained based on paired CT images and corresponding ROIs to make fine adjustments to the coarse segmentation results.

  \item ($9^{th}$ place, Y. Zhang \textit{et al.}) This method was based on nnUNet~\citep{isensee2021nnu} framework. The spacing of CT images and corresponding labels were resampled to [3, 1, 1]. To discard the background, the data and corresponding label were cropped based on the body bounding box. Data augmentation methods, including spatial-, intensity- and label-based transformation, were used to enhance data diversity and richness. Paired CT images were randomly cropped into patches of size [28, 224, 224] and used to train a 3D full-resolution UNet based on nnUNet~\citep{isensee2021nnu}. In the inference stage, the patch size was equal to the patch size during training, and the neighbourhood inference was performed using intervals of 1/2 patch size.

  \item ($10^{th}$ place, C. Lee \textit{et al.}) This method had two successive stages: localization and segmentation. In the localization stage, a bounding box was identified to encompass the OARs utilizing a 2D-based object detection network powered by the YOLO-v7 model~\citep{wang2022efficient}. In the segmentation stage, different window widths and levels were used for multi-channel input generation. A segmentation network with DynUNet architecture was trained with these multi-channel inputs to enhance the ability to distinguish detailed features. In the inference stage, ROIs were first extracted, and the segmentation network was then used to generate the final predictions.
  
  \item ($11^{th}$ place, K. Huang \textit{et al.}) The proposed method was based on nnUNetV2~\citep{isensee2021nnu} framework, which is the same as that they used for Task01. To adapt the model for GTV segmentation, the parameters were set following~\citep{isensee2021nnu}. In the inference stage, various forms of augmentation were applied, such as different region cropping and adjustments in scaling. The average of predictions was taken as the final results.

\end{itemize}

\begin{table}[t]
  \centering
  \caption{Summary of statistical significance analysis ($p$-value) for the top 3 teams on the OAR segmentation task.}
  \resizebox{0.9\columnwidth}{!}{
    \begin{tabular}{lccc|ccc}
    \hline
    \multirow{2}[4]{*}{Team} & \multicolumn{3}{c|}{DSC} & \multicolumn{3}{c}{NSD} \bigstrut\\
\cline{2-7}          & Y. Zhong \textit{et al.} & Y. Ye \textit{et al.} & Y. Su \textit{et al.} & Y. Zhong \textit{et al.} & Y. Ye \textit{et al.} & Y. Su \textit{et al.} \bigstrut\\
    \hline
    Brain & 0.19  & 0.46  & 0.19  & 0.54  & 0.94  & 0.53 \bigstrut[t]\\
    BrainStem & 0.10  & 0.04  & 0.61  & 0.18  & 0.08  & 0.62 \\
    Chiasm & 0.49  & 0.07  & 0.80  & 0.59  & 0.07  & 0.78 \\
    Cochlea\_L & 0.10  & 0.54  & 0.23  & 0.23  & 0.51  & 0.11 \\
    Cochlea\_R & 8.99e-04 & 0.60  & 0.20  & 3.06e-04 & 0.56  & 0.24 \\
    Esophagus & 0.03  & 0.04  & 0.08  & 0.23  & 0.09  & 0.21 \\
    ETbone\_L & 0.01  & 0.04  & 0.14  & 0.02  & 0.04  & 0.07 \\
    ETbone\_R & 0.22  & 0.55  & 0.39  & 0.37  & 0.90  & 0.88 \\
    Eye\_L & 0.18  & 0.66  & 0.26  & 0.30  & 0.63  & 0.19 \\
    Eye\_R & 0.57  & 0.80  & 0.61  & 0.70  & 0.69  & 0.76 \\
    Hippocampus\_L & 0.53  & 0.74  & 0.77  & 0.77  & 0.96  & 0.64 \\
    Hippocampus\_R & 0.02  & 0.25  & 0.78  & 0.02  & 0.34  & 0.75 \\
    IAC\_L & 0.62  & 0.03  & 0.42  & 0.94  & 0.05  & 0.48 \\
    IAC\_R & 3.67e-07 & 0.86  & 0.16  & 2.27e-07 & 0.75  & 0.18 \\
    Larynx & 3.84E-11 & 0.25  & 0.07  & 3.32e-06 & 0.21  & 0.11 \\
    Larynx\_Glottic & 0.18  & 0.07  & 0.10  & 0.90  & 0.23  & 0.13 \\
    Larynx\_Supraglot & 0.03  & 0.16  & 0.13  & 0.21  & 0.52  & 0.23 \\
    Lens\_L & 0.14  & 0.21  & 0.81  & 0.21  & 0.95  & 0.21 \\
    Lens\_R & 0.13  & 0.12  & 0.30  & 0.11  & 0.46  & 0.64 \\
    Mandible\_L & 0.24  & 0.07  & 0.90  & 0.94  & 0.45  & 0.79 \\
    Mandible\_R & 0.34  & 8.23e-05 & 3.56e-03 & 0.72  & 0.50  & 0.41 \\
    Mastoid\_L & 0.26  & 0.33  & 0.39  & 0.37  & 0.64  & 0.21 \\
    Mastoid\_R & 0.21  & 0.04  & 0.44  & 0.32  & 0.69  & 0.30 \\
    MiddleEar\_L & 0.80  & 0.16  & 0.40  & 0.77  & 0.82  & 0.25 \\
    MiddleEar\_R & 4.97e-06 & 4.35e-06 & 1.89e-04 & 9.61e-03 & 5.04e-06 & 4.26e-04 \\
    OpticNerve\_L & 0.54  & 0.20  & 0.38  & 0.30  & 0.36  & 0.94 \\
    OpticNerve\_R & 0.13  & 0.82  & 0.39  & 0.11  & 0.68  & 0.19 \\
    OralCavity & 6.54e-02 & 0.29  & 3.97e-08 & 0.08  & 0.51  & 7.31e-09 \\
    Parotid\_L & 0.02  & 0.65  & 7.64e-05 & 0.06  & 0.51  & 2.70E-10 \\
    Parotid\_R & 0.34 & 0.74  & 0.13  & 0.86  & 0.22  & 0.08 \\
    PharynxConst & 0.60  & 0.27  & 0.20  & 0.74  & 0.32  & 0.11 \\
    Pituitary & 0.89  & 0.54  & 0.42  & 0.96  & 0.39  & 0.38 \\
    SpinalCord & 0.07  & 0.68  & 0.06  & 0.11  & 0.40  & 0.23 \\
    Submandibular\_L & 0.18  & 0.66  & 0.02  & 0.10  & 0.54  & 0.05 \\
    Submandibular\_R & 0.71  & 0.03  & 0.68  & 0.33  & 0.06  & 0.95 \\
    TemporalLobe\_L & 0.18  & 0.35  & 0.44  & 0.35  & 0.35  & 0.49 \\
    TemporalLobe\_R & 0.09  & 0.49  & 0.35  & 0.11  & 0.55  & 0.35 \\
    Thyroid & 0.17  & 0.33  & 0.08  & 0.33  & 0.13  & 0.17 \\
    Trachea & 2.74e-05 & 0.50  & 0.92  & 1.73e-08 & 0.70  & 1.00 \\
    TympanicCavity\_L & 5.19e-03 & 0.08  & 3.52e-03 & 0.04  & 0.55  & 1.98e-03 \\
    TMjoint\_L & 0.81  & 0.91  & 0.34  & 0.87  & 0.45  & 0.97 \\
    TMjoint\_R & 4.15e-03 & 0.66  & 0.29  & 0.14  & 0.98  & 0.16 \\
    TympanicCavity\_R & 4.48e-05 & 0.35  & 0.04  & 0.06  & 0.61  & 2.63e-03 \\
    VestibulSemi\_L & 1.76e-03 & 0.96  & 0.02  & 8.34e-03 & 0.39  & 0.02 \\
    VestibulSemi\_R & 0.30  & 0.16  & 0.36  & 0.35  & 0.37  & 0.34 \bigstrut[b]\\
    \hline
    Average & 1.18e-06 & 0.08  & 0.15  & 2.21e-05 & 0.88  & 0.03 \bigstrut\\
    \hline
    \end{tabular}%
    }
  \label{tab:task01_ttest}%
\end{table}%

\begin{table}[t]
  \centering
  \caption{Rankings of methods in DSC and NSD scores for GTV segmentation}
    \scalebox{0.6}{
    \begin{tabu}{cccc|cccc}
    \hline
    \multirow{2}[3]{*}{Method} & \multicolumn{3}{c|}{DSC Rank} & \multicolumn{3}{c}{NSD Rank} & \multirow{2}[3]{*}{Overall} \bigstrut\\
    \cline{2-7}          & GTVp  & GTVnd & Average & GTVp  & GTVnd & Average &  \bigstrut[t]\\
    \hline
    M. Astaraki \textit{et al.} & 3     & 4     & 3.5   & 1     & 4     & 2.5   & 1 \\
    Y. Ye \textit{et al.} & 2     & 3     & 2.5   & 2     & 6     & 4     & 2 \\
    Z. Xing \textit{et al.} & 7     & 1     & 4     & 3     & 2     & 2.5   & 3 \\
    K. Yang \textit{et al.} & 1     & 5     & 3     & 4     & 5     & 4.5   & 4 \\
    C. Ulrich \textit{et al.} & 8     & 2     & 5     & 6     & 1     & 3.5   & 5 \\
    N. Ndipenoch \textit{et al.} & 5     & 6     & 5.5   & 5     & 3     & 4     & 6 \\
    Y. Su \textit{et al.} & 6     & 7     & 6.5   & 7     & 7     & 7     & 7 \\
    J. Huang \textit{et al.} & 4     & 8     & 6     & 8     & 8     & 8     & 8 \\
    Y. Zhang \textit{et al.} & 10    & 9     & 9.5   & 9     & 9     & 9     & 9 \\
    C. Lee \textit{et al.} & 9     & 11    & 10    & 10    & 11    & 10.5  & 10 \\
    K. Huang \textit{et al.} & 11    & 10    & 10.5  & 11    & 10    & 10.5  & 11 \bigstrut[b]\\
    \hline
    \end{tabu}}%
  \label{tab:task02_rank}%
\end{table}%

\begin{table}[t]
  \centering
  \caption{Summary of the quantitative evaluation results of GTVp and GTVnd segmentation by the eleven teams.}
  \setlength\tabcolsep{4pt}
    \scalebox{0.5}{\begin{tabu}{lccc|ccc}
    \hline
    \multirow{2}[4]{*}{Team} & \multicolumn{3}{c|}{DSC (\%)} & \multicolumn{3}{c}{NSD (\%)} \bigstrut\\
\cline{2-7}          & GTVp  & GTVnd & Average & GTVp  & GTVnd & Average \bigstrut\\
    \hline
    M. Astaraki \textit{et al.} & 78.56±7.54 & 67.75±14.64 & 73.15±12.83 & \textbf{36.61±12.17} & 63.15±16.24 & 49.88±19.55 \bigstrut[t]\\
    Y. Ye \textit{et al.} & 78.76±7.16 & 68.10±12.17 & 73.43±11.31 & 36.45±11.70 & 62.26±15.57 & 49.36±18.87 \\
    Z. Xing \textit{et al.} & 78.07±7.82 & \textbf{69.28±12.12} & \textbf{73.68±11.11} & 36.44±12.25 & 64.04±14.37 & \textbf{50.24±19.20} \\
    K. Yang \textit{et al.} & \textbf{78.76±6.60} & 67.41±13.78 & 73.09±12.21 & 35.92±11.05 & 63.08±15.37 & 49.50±19.07 \\
    C. Ulrich \textit{et al.} & 77.71±7.79 & 69.18±12.80 & 73.44±11.42 & 35.60±11.66 & \textbf{64.76±15.04} & 50.18±19.84 \\
    N. Ndipenoch \textit{et al.} & 78.25±7.54 & 67.21±14.52 & 72.73±12.82 & 35.90±11.87 & 63.31±15.78 & 49.61±19.56 \\
    Y. Su \textit{et al.} & 78.13±7.27 & 66.91±14.54 & 72.52±12.79 & 35.21±11.11 & 62.24±16.00 & 48.73±19.30 \\
    J. Huang \textit{et al.} & 78.36±7.09 & 66.36±14.09 & 72.36±12.66 & 34.18±10.26 & 61.96±15.48 & 48.07±19.12 \\
    Y. Zhang \textit{et al.} & 76.89±7.37 & 66.25±12.74 & 71.57±11.69 & 33.22±10.66 & 60.30±13.94 & 46.76±18.37 \\
    C. Lee \textit{et al.} & 77.46±7.53 & 63.39±13.85 & 70.42±13.18 & 32.96±10.69 & 55.62±14.51 & 44.29±17.05 \\
    K. Huang \textit{et al.} & 76.71±6.85 & 65.97±12.04 & 71.34±11.17 & 32.76±9.61 & 59.70±13.34 & 46.23±17.79 \bigstrut[b]\\
    \hline
    Baseline & 75.80±7.28 & 66.83±11.48 & 71.32±10.61 & 33.41±11.61 & 61.49±13.06 & 47.45±18.70 \bigstrut\\
    \hline
    \end{tabu}
    }
  \label{tab:task02_dsc_nsd}%
\end{table}%

\begin{table}[t]
  \centering
  \caption{Summary of statistical significance analysis ($p$-value) for the top 3 teams on the GTV segmentation task.}
  \resizebox{\columnwidth}{!}{
    \begin{tabular}{lccc|ccc}
    \hline
    \multirow{2}[4]{*}{Team} & \multicolumn{3}{c|}{DSC} & \multicolumn{3}{c}{NSD} \bigstrut\\
\cline{2-7}          & M. Astaraki \textit{et al.} & Y. Ye \textit{et al.} & Z. Xing \textit{et al.} & M. Astaraki \textit{et al.} & Y. Ye \textit{et al.} & Z. Xing \textit{et al.} \bigstrut\\
    \hline
    GTVp  & 0.55  & 0.16  & 0.18  & 0.81  & 0.99  & 0.54 \bigstrut[t]\\
    GTVnd & 0.68  & 0.17  & 0.12  & 0.30  & 0.04 & 0.34 \bigstrut[b]\\
    \hline
    Average & 0.55  & 0.60  & 0.41  & 0.38  & 0.13  & 0.32 \bigstrut\\
    \hline
    \end{tabular}%
    }
  \label{tab:task02_ttest}%
\end{table}%

\begin{figure*}[t]
    \centering
    \includegraphics[width=0.94\textwidth]{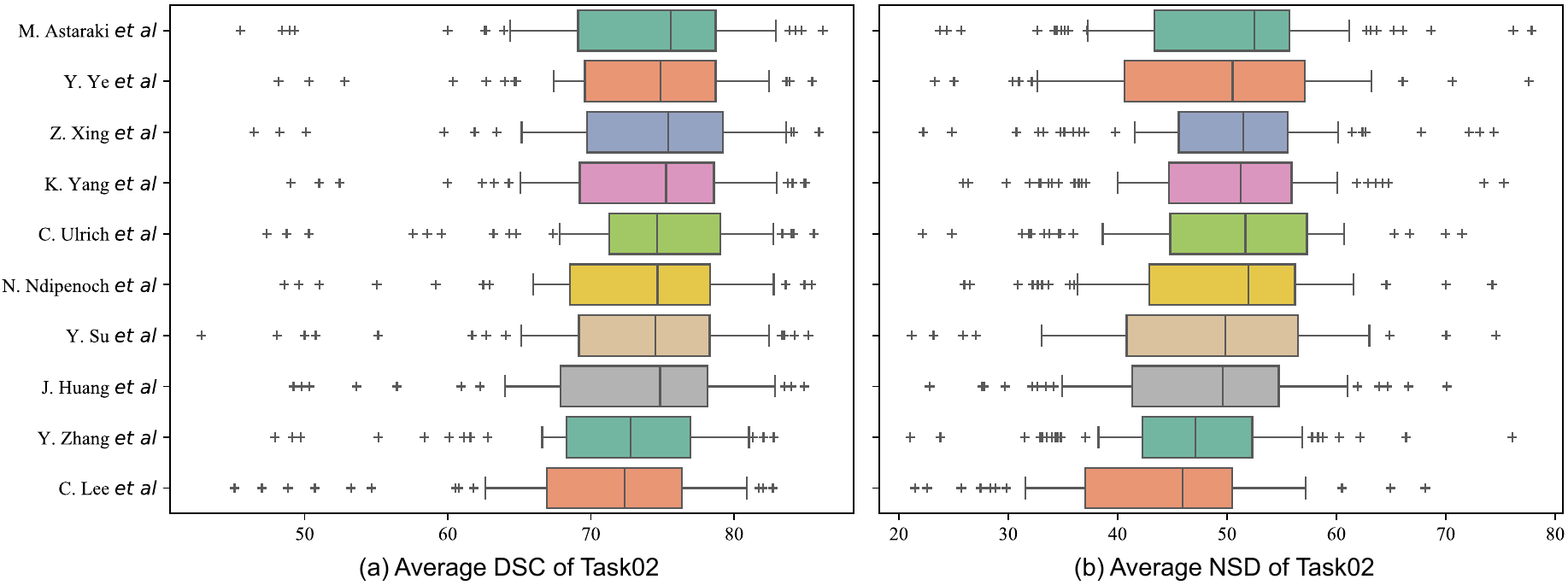}
    \caption{Box plot of the patient-level average segmentation performance for GTVs in terms of DSC and NSD.}
    \label{fig:task02_avg_results}
\end{figure*}

\begin{figure}[t]
    \centering
    \includegraphics[width=0.94\columnwidth]{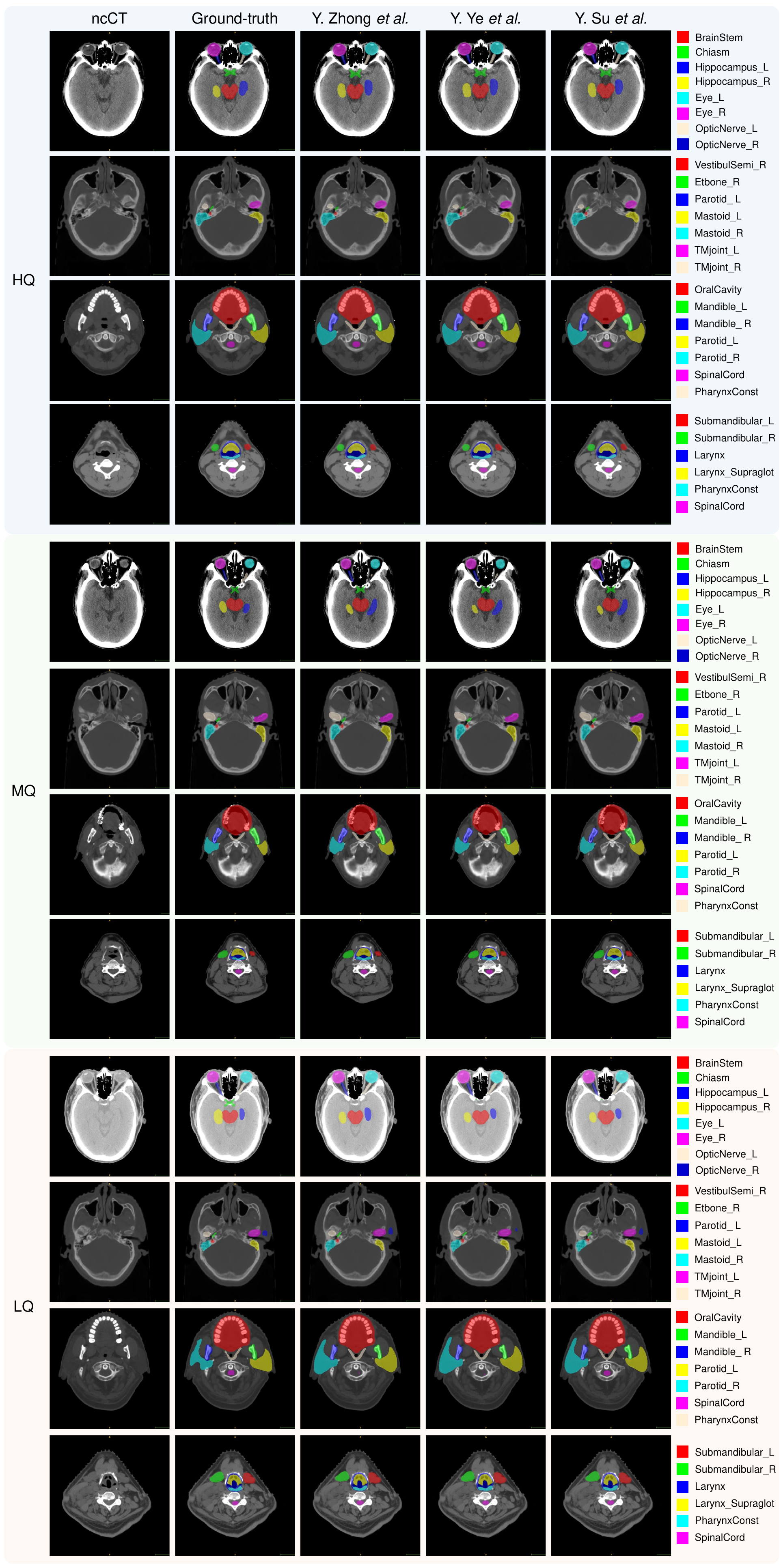}
    \caption{Qualitative OAR segmentation using the Top3 teams on the SegRap2023 testing set.}
    \label{fig:task01_vis}
\end{figure}

\begin{figure}[t]
    \centering
    \includegraphics[width=0.94\columnwidth]{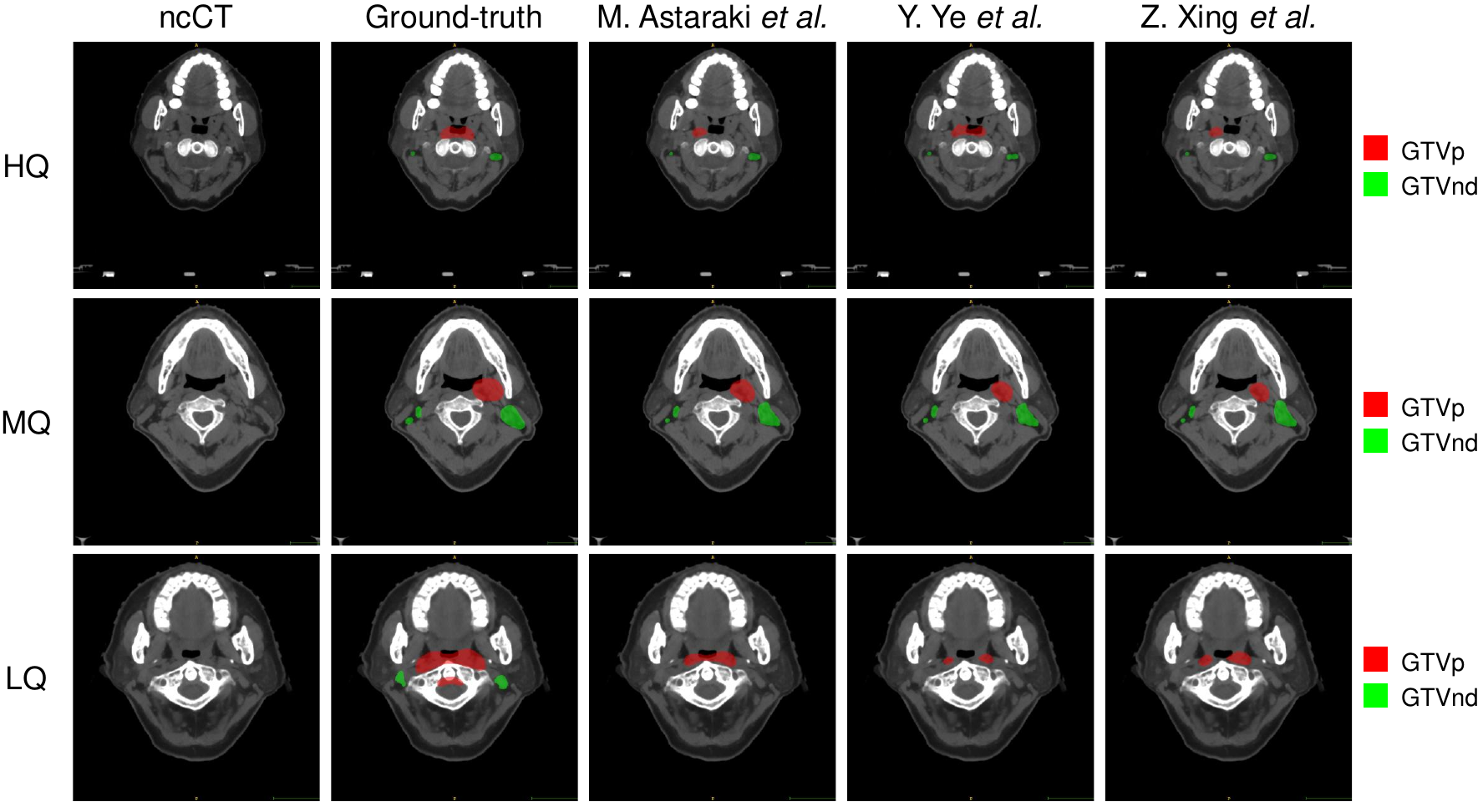}
    \caption{Qualitative GTV segmentation using the Top3 teams on the SegRap2023 testing set.}
    \label{fig:task02_vis}
\end{figure}

\section{Results}\label{sec:set5}
\subsection{Results of the Task01}
The final ranking results of Task01 are listed in Table~\ref{tab:task1_rank} sorted by their scores. Table~\ref{tab:task01_DSC} and Table~\ref{tab:task01_NSD} present the detailed performance of each OAR in each team in terms of DSC and NSD, respectively. It can be observed that the winner (Y. Zhong \textit{et al.}) achieved the best performance on more than 30 OARs and most of the rest of the OARs performances also ranked in the top 3. The top 3 teams achieved promising performance whose an average score of over 86\% in terms of DSC and NSD. The other four teams also obtained promising results with averages DSC and NSD larger than 80\%. However, it does not mean that the OARs segmentation is completely solved because these methods still perform poorly on some small, complex organs (DSC/NSD $<$ 80\%).

\par Interestingly, almost all teams used the nnUNet~\citep{isensee2021nnu} or its variants as the baseline, but their performances were hugely different. For example, the performance of the winners and the K. Huang \textit{et al.} methods is significantly different, 86.70\% \textit{vs} 78.14\% in terms of DSC score. Meanwhile, we also evaluated the baseline pure nnUNet (with the default setting of 3d\_fullres) in the official tutorial and listed the results in the last columns of Table~\ref{tab:task01_DSC} and Table~\ref{tab:task01_NSD}. It can be found that there are four teams that performed worse than the baseline. This highlights the necessity of designing some specified data-processing strategies, network modules or training or testing approaches for this task according to the data characteristics.

\par We calculated the paired \textit{t-test} between the ranking \textit{n-th} team and the ranking \textit{(n+1)-th} team (\textit{n} ranges from 1 to 3). Table~\ref{tab:task01_ttest} presents the statistical analysis results of the top 3 teams. It can be observed that the winner is significantly superior to the second place in terms of average DSC and NSD. However, there are no significant differences between the second and third teams except for the absolute values of DSC and NSD. Compared with the fourth team, the third team achieved significantly better NSD scores and comparable DSC scores. The above results show the final leaderboard is robust and convincing.

\subsection{Results of Task02}
Table~\ref{tab:task02_rank} presents the final ranking scores of the GTV segmentation. The detailed performances of each team are shown in Table~\ref{tab:task02_dsc_nsd} and Fig.~\ref{fig:task02_avg_results}. M. Astaraki \textit{et al.} won first place with the average ranking score of 3. Y. Ye \textit{et al.} and Z. Xing \textit{et al.} achieved the same average ranking score of 3.25, but the standard deviation of Y. Ye \textit{et al.} was smaller, so the final ranking results were that Y. Ye \textit{et al.} and Z. Xing \textit{et al.} won the second and third places, respectively. There are four teams that obtained encouraging performance with greater than 73\% mean DSC scores. In addition, all submissions of Task02 performed well on the GTVp segmentation with larger than 76.71\% DSC, and the score of DSC in GTVnd segmentation has a larger variability ranging from 63.39\% to 69.28\%. In addition, we also found that most of the methods can not achieve promising performances on both GTVp and GTVnd segmentation at the same time. These results demonstrated that the automatic GTVp and GTVs contouring is still a challenging and unsolved problem, and more attention should be paid to research to improve the segmentation performance further.

\par Different from the results of Task01, these teams that used the nnUNet or its variants achieved similar results on the GTV segmentation task. The average performance gap between the winner and the 11-$th$ ranking team was nearly 2\% and 3\% in terms of DSC and NSD. Compared with the pure nnUNet baseline (the last line in Table~\ref{tab:task02_dsc_nsd}), eight teams achieved better results in both terms of DSC and NSD. Although the segmentation results are consistent and robust, there are huge performance gaps between these methods and real clinical requirements. 

\par Table~\ref{tab:task02_ttest} presents a detailed statistical analysis of the top 3 teams. The results show that there are no significant performance differences in terms of DSC and NSD between the winner and the second-place method except for the numerical values and the ranking scores. Similar trends can be found in the pair of the second, and third places, no significant performance differences were found except for the NSD score in GTVnd segmentation. Besides, from Table~\ref{tab:task02_dsc_nsd} and Table~\ref{tab:task02_ttest}, it can be noticed that Z. Xing \textit{et al}  obtained the best average performance in both terms of DSC and NSD, but this team ranked on the third place due to the poor overall ranking score. In addition, C. Ulrich \textit{et al}  achieved the best NSD and second DSC in GTVnd segmentation and were not even included in the top 3 teams yet caused by the insufficient results in GTVp segmentation.  These results show the ranking scheme of this challenge (rank-then-aggregate~\citep{dorent2023crossmoda}) is robust and alleviates the impact of some extremely good or bad results.



\subsection{Visualization}
Fig.~\ref{fig:task01_vis} visually presents the OAR segmentation outcomes from the top three performing teams. To show segmentation differences, we selected three patients based on the lower quartile (LQ), median quartile (MQ), and high quartile (HQ) of the average across both the top three teams and the 45 OARs. The results highlight that these methods excel in achieving accurate segmentations for larger organs such as BrainStem, Parotid$\_$L, and Parotid$\_$R. However, challenges persist in accurately segmenting small and intricate organs. For instance, the Chiasm exhibits under-segmentation, particularly in the case of the LQ patient. Fig.~\ref{fig:task02_vis} visualizes the GTV segmentation results of the top 3 teams. These results show that the GTVp and GTVnd segmentation are still challenging. Specifically, most GTVp segmentation results suffer from under-segmentation (in HQ, MQ and LQ patients). Additionally, some GTVnd even can not be identified and segmented (in the LQ patient). These findings highlight the challenge of achieving precise and automated GTV segmentation, which warrants heightened attention and further investigation.

\section{Discussion}\label{sec:set6}
In this section, we discuss the potential solutions, limitations and future direction of automatic segmentation in radiation therapy planning and provide some insights about the clinically applicable OAR and GTV segmentation.
\subsection{OAR segmentation in head and neck}
All submitted algorithms demonstrated that supervised learning can achieve promising mean performance ($>$ 80\%) in terms of DSC and NSD. However, the results of some complex OARs are still not good enough ($<$ 80\%). The reason may be most of these solutions are based on one-stage segmentation and do not apply specific designs for complex or small organs. The winner's solution demonstrated the specifically designed structure-specific label generation and boundary refinement can obtain encouraging performance improvement over the baseline. Meanwhile, there are imbalance problems and inequality optimization when segmenting 45 OARs directly. Applying the balance loss~\citep{lin2017focal} and stratified optimization~\citep{ye2022comprehensive} may bring benefits to improve the segmentation performance of the small and complex OAR, but there are no participants that have investigated the performance of these methods. 

\par Recently, the universal model with transfer learning has shown promising performance on multiple medical image segmentation tasks~\citep{liu2023clip,Ye2023UniSeg,wang2023mis}. The second-place solution shows the transferable ability of the universal model~\citep{Ye2023UniSeg} from other tasks to the head and neck OAR segmentation. The third-place method proved that using the large patch size and not applying mirror spatial augmentation also can boost the baseline performance, suggesting that the simple task-driven data processing methods can lead to benefits. Note that the top 3 teams reached a promising performance that is superior to previous head and neck OAR segmentation studies although with different datasets. These results also provided a fair baseline and benchmarking results for further research.

\subsection{GTV segmentation of NPC}
All submitted methods for GTV segmentation obtained comparable results, however, no team surpassed 80\% in terms of DSC or NSD. The top 3 teams applied the two-stage segmentation with intensity distribution harmonization, transfer learning and test time augmentation strategies to handle the inherent and challenging problems in GTV segmentation, respectively. However, the segmentation results of the top 3 teams have under-segmentation and even targets missing, as shown in Fig.~\ref{fig:task02_vis}. In addition, the segmentation performance gaps between this challenge and previous works are huge and different~\citep{luo2023deep,li2022npcnet,liao2022automatic,lin2019deep}. These results highlight the urgency of developing an accurate GTV segmentation method to handle the inherent challenges and further evaluate in the clinical practice. 

\par There are some potential directions to enhance the GTV segmentation performance: 1) exploiting the position and boundary-aware feature attention to describing the variable location and irregular boundary of GTV~\citep{li2022npcnet}; 2) investigating the performance improvement by using the OAR segmentation to provide the anatomical information~\citep{yan2023anatomy}; 3) mining the complementary information across ncCT and ceCT scans to highlight the target representation, which not be noticed by recent works.

\subsection{Clinical application}
The ultimate goal of developing automatic OAR and GTV segmentation methods is to accelerate the clinical delineation workflow and reduce the radiation oncologists' burden. There are several challenges to fulfilling the above-mentioned purposes. Firstly, accurate segmentation is the most important criterion for precision radiotherapy. Precise delineation of GTVs is essential to ensure accurate delivery of radiation dose to the affected region, thereby enhancing treatment effectiveness. Accurate OAR segmentation can lead to minimizing the received radiation dose during the treatment process to reduce treatment-induced side effects. Secondly, robust and generalizable segmentation across different hospitals, scanners, imaging protocols and patients is desirable. The domain shift may lead to performance degradation significantly and further cause the results to be not clinically acceptable. Thirdly, efficient and easily editable is still important to build a user-friendly delineation system. In clinical practice, quick response is necessary as most automatic segmentation can not be applied in clinical directly and needs radiation oncologists to refine, especially for the online intensity-modulated radiation therapy system. \cite{tang2019clinically} claimed that the deep learning-based automatic contouring system with a mean DSC of 78.34\% over 28 OARs was clinically applicable after minor revision. According to this study, most participants achieved clinically applicable results for most OARs. \cite{liao2022automatic} and~\cite{luo2023deep} have performed clinical studies on GTVp and GTVnd segmentation and shown that the deep learning segmentation system can be clinically accepted with few refinements when the DSC of GTVp and GTVnd are greater than 83\% and  80\%. So, there are huge gaps between the performance of these participants and the clinically acceptable results for the GTVs.

\subsection{Limitation and future direction}
Compared with the abdominal organ and tumor segmentation~\citep{luo2021word,ma2021abdomenct}, there are very few works that have built large-scale datasets and comprehensively evaluated the performance of recent methods for the OARs and GTVs of head and neck cancer. Although this work has developed a large-scale dataset and evaluated more than ten cut-edge methods,  it still faces limitations in terms of robustness and generalization evaluation, primarily attributed to the absence of a multi-center dataset. Additionally, the dataset exclusively focuses on NPC patients, overlooking the diverse range of patients encompassed by head and neck cancer. Despite the inclusion of annotations for 45 OARs and 2 GTVs in the SegRap2023 challenge, there is an omission of several radiotherapy-required clinically target volumes (CTV). To address these shortcomings, we plan to enlarge the scale of the dataset and data source and further extend the segmentation tasks to more categories in the next year.

\section{Conclusion}\label{sec:set7}
This work summarises the results of all participants in the SegRap2023 challenge. The challenge provided 200 paired CT scans for OAR and GTV segmentation model development and evaluation. To the best of our knowledge, SegRap2023 is the most comprehensive and exhausted labelled dataset to evaluate the OAR and GTV segmentation. Ten and eleven teams submitted their solutions for benchmarking and comparison for OAR and GTV segmentation, respectively, which have been introduced and analyzed in detail.  The results show that most large-size OARs can be segmented accurately and can be seen as a well-solved problem. However, for the small-size OARs and GTVs, there are huge gaps between the performance of participants and the clinically acceptable, suggesting that future research should focus on these unsolved problems more. In the future, we plan to extend this challenge in the aspect of data scale, source and categories to be more suitable for the clinical requirement.
\section{Acknowledgment}
This work was supported by the National Natural Science Foundation of China [Grant 62271115, Grant 82203197], the Sichuan Science and Technology Program, China (Grant 2022YFSY0055, 2023NSFSC1852), the Sichuan Provincial Cadre Health Research Project (Grant/award number: 2023-803) and the Radiation Oncology Key Laboratory of Sichuan Province Open Fund (2022ROKF04). We would like to thank M.D. S.C. Zhang and M.D. W. Liao and their team members for data collection, annotation, and checking. We also would like to thank the support team of the Grand Challenge Platform and the MICCAI challenge organization team for their sincere help while hosting the challenge. We also would like to thank all participants for their active participation and working hard.

\bibliographystyle{model2-names.bst}\biboptions{authoryear}
\bibliography{refs}

\end{document}